\documentclass[12pt]{article}
\usepackage[latin9]{inputenc}
\usepackage[a4paper]{geometry}
\usepackage{float}
\usepackage{amsmath}
\usepackage{amssymb}
\usepackage{amsthm}
\usepackage{graphicx}
\usepackage{bm,booktabs}
\usepackage{hyperref}

\newtheorem{theorem}{Theorem}[section]
\newtheorem{proposition}[theorem]{Proposition}
\newtheorem{lemma}[theorem]{Lemma}
\newtheorem{corollary}[theorem]{Corollary}

\newtheorem{assumption}{Assumption}

\theoremstyle{definition}
\newtheorem{example}{Example}

\theoremstyle{remark}
\newtheorem{remark}{Remark}

\usepackage[authoryear]{natbib}

\makeatletter

\date{November, 2026}

\makeatother

\begin{document}
\title{Comparing Two-Sample IV Estimators under Weak Instruments}
\author{Jeffrey Tse\thanks{Email: jt801@cam.ac.uk} \\
 Department of Statistics, University of Oxford, United Kingdom \\
 Faculty of Economics, University of Cambridge, United Kingdom}
\maketitle
\begin{abstract}
\baselineskip=13pt

\noindent We compare three estimators in a two-sample summary-data instrumental variables (IV) model: inverse-variance weighting (IVW), debiased inverse-variance weighting (dIVW), and the profile score (PS) estimator. The framework assumes independent samples and instrument-specific summary estimates, with two-sample Mendelian randomisation as a leading application. We compare existing fixed- and many-weak-instrument results and the sufficient conditions for the three estimators. Under the variance scaling used in the profile-score literature, the principal collective-strength conditions for PS and dIVW are equivalent, although their additional restrictions on instrument concentration and growth differ. We then compare their asymptotic variance formulae for fixed and diverging numbers of instruments. For fixed $L$, IVW and dIVW have the same first-order asymptotic variance. Under a common variance-ratio condition, the PS variance formula exceeds the IVW variance by a positive term of order $n^{-2}$, compared with the order-$n^{-1}$ leading variance. For diverging $L$, we establish variance orders across instrument-strength regimes and the exact ordering $\mathsf{V}_{\mathtt{IVW}} < V_{\mathtt{PS}} \leq \mathsf{V}_{\mathtt{dIVW}}$ under the same variance-ratio condition. Absolute variance gaps vanish when $\sqrt{L}\kappa\to\infty$, whereas relative equivalence holds if and only if $\kappa\to\infty$. These comparisons concern asymptotic variance formulae rather than finite-sample efficiency or mean squared error.
\end{abstract}


\thispagestyle{empty}

\noindent \baselineskip=20pt

\pagebreak{}

\pagenumbering{arabic} \setcounter{page}{1}

\section{Introduction}
Two-sample instrumental variables (IV) methods are useful when the treatment or exposure and the outcome are observed in separate samples, while a common set of instruments is available in both. The first sample identifies instrument--treatment associations and the second identifies instrument--outcome, or reduced-form, associations. This design arises in several applications and is especially prominent in two-sample Mendelian randomisation (MR), where genetic variants serve as instruments and genome-wide association studies provide summary association estimates \citep{10.1093/aje/kwt084}.

This paper studies a summary-data two-sample IV model in which the instrument-specific first-stage and reduced-form estimates are independent across instruments. This diagonal covariance structure is natural, for example, in MR after instruments have been pruned or clumped to be approximately uncorrelated. It is more restrictive than a general two-sample IV model with correlated instruments and full covariance matrices; our results should therefore be interpreted as theory for the independent-summary-association setting rather than unrestricted two-sample IV.

We compare IVW, dIVW, and PS estimators. Two-sample IV estimation has a longstanding econometric literature \citep{10.1162/REST_a_00011,PACINI201650}, while IVW, dIVW, and PS have been developed and studied extensively in the summary-data MR literature \citep{10.1214/19-AOS1866,10.1214/20-AOS2027}. We place these results in a common summary-data IV framework and compare the sufficient conditions required by the three estimators. As a preliminary step, we establish fixed-$L$ consistency and asymptotic normality for IVW and dIVW directly within the common summary-data framework, providing the variance expressions needed for our subsequent comparisons. Our first contribution is to relate their sufficient conditions under many weak instruments: under the variance scaling used for PS, its collective-strength condition is equivalent to the corresponding dIVW condition, while the two procedures differ in their additional restrictions on the concentration of instrument strength and the growth of the instrument set. Our second contribution is to compare the asymptotic variance formulae of IVW, dIVW, and PS under both fixed and diverging numbers of instruments. For fixed numbers of instruments, under a common variance-ratio condition, the leading PS variance component coincides exactly with the IVW variance, whereas the remaining positive component is associated with profile-score information bias. This additional component is of smaller order than the leading variance under fixed-$L$ asymptotics, so PS and IVW/dIVW are first-order asymptotically equivalent despite the higher-order variance difference. For diverging numbers of instruments, we establish the variance orders across different instrument-strength regimes and show that, under the same variance-ratio condition, the IVW variance formula is strictly smaller than the PS variance formula, which is no greater than the dIVW variance formula. We further show that the absolute variance gaps vanish when $\sqrt{L}\kappa\to\infty$, whereas relative equivalence of all three variance formulae holds if and only if $\kappa\to\infty$. These comparisons concern the asymptotic variance formulae and do not, by themselves, establish an ordering of finite-sample efficiency or mean squared error. The arguments depend on the two-sample summary-data model and valid-IV proportionality, rather than on the instruments being genetic variants. MR provides the principal motivating example, but the theoretical results apply to any two-sample IV setting satisfying the stated assumptions.

\section{Two-Sample Summary-Data IV Framework}
Let $\beta$ be the structural effect of a treatment or exposure $X$ on an outcome $Y$, and let $Z_1,\ldots,Z_L$ denote valid instruments. We work directly with the first-stage and reduced-form coefficients. Let $\gamma_j$ denote the association of instrument $j$ with $X$ and $\Gamma_j$ its association with $Y$. Validity and a common linear structural effect imply the proportionality restriction
\begin{assumption}[Valid-IV proportionality]\label{assume:structure}
For all $j=1,\ldots,L$, $\Gamma_j=\beta\gamma_j$. Additionally, the $\gamma_j$'s are not all zero.
\end{assumption}
This restriction identifies $\beta$ from the first-stage and reduced-form coefficients. It is the summary-data analogue of the familiar reduced-form restriction in a linear IV model.

We consider two distinct, independent samples. Let $n_X$ and $n_Y$ be the treatment and outcome sample sizes. From the treatment sample we obtain estimates $\hat\gamma_j$ of the first-stage coefficients $\gamma_j$; from the outcome sample we obtain estimates $\hat\Gamma_j$ of the reduced-form coefficients $\Gamma_j$. In the simplest implementation these are marginal regression coefficients,
\begin{align*}
X_i &= \gamma_{0,j}+\gamma_j Z_{ij}+e_{X,ij}, \qquad i=1,\dots,n_X,\\
Y_i &= \Gamma_{0,j}+\Gamma_j Z_{ij}+e_{Y,ij}, \qquad i=1,\dots,n_Y.
\end{align*}
We assume $n_X/n_Y\to a$, where $0<a<\infty$, as $n_X+n_Y\to\infty$.

For fixed $L$, suppose the summary estimators satisfy
\begin{align*}
\sqrt{n_Y}(\hat\Gamma_j-\Gamma_j)&\overset{d}{\longrightarrow}\mathcal{N}(0,\sigma_{\Gamma_j}^2),\\
\sqrt{n_X}(\hat\gamma_j-\gamma_j)&\overset{d}{\longrightarrow}\mathcal{N}(0,\sigma_{\gamma_j}^2),
\end{align*}
with variance estimators satisfying $n_Y\hat\sigma_{Y_j}^2\overset{p}{\to}\sigma_{\Gamma_j}^2$ and $n_X\hat\sigma_{X_j}^2\overset{p}{\to}\sigma_{\gamma_j}^2$. Define $\sigma_{Y_j}^2=\sigma_{\Gamma_j}^2/n_Y$ and $\sigma_{X_j}^2=\sigma_{\gamma_j}^2/n_X$.

For the many-instrument theory we use the following stronger working model.
\begin{assumption}[Independent summary-association model]\label{assume:measure-error}
(1) The pairs $\{(\hat\Gamma_j,\hat\gamma_j)\}_{j=1}^L$ are mutually independent.\\
(2) For all $j$,
$$
\begin{pmatrix}\hat\Gamma_j\\ \hat\gamma_j\end{pmatrix}
\sim \mathcal N\!\left(
\begin{pmatrix}\Gamma_j\\ \gamma_j\end{pmatrix},
\begin{pmatrix}\sigma_{Y_j}^2&0\\0&\sigma_{X_j}^2\end{pmatrix}\right),
$$
where the ratios $\sigma_{Y_j}^2/\sigma_{X_j}^2$ are uniformly bounded above and bounded away from zero.
\end{assumption}
The within-instrument independence follows from the independence of the two samples. Independence across instruments is an additional restriction: it holds directly for independent instruments under the corresponding marginal summary regressions and is a useful approximation in two-sample MR after LD pruning or clumping. General two-sample IV with correlated instruments instead requires the full covariance matrices of the first-stage and reduced-form coefficient vectors and is outside the scope of the present results.

A canonical application is two-sample MR. There, $Z_j$ is a genetic variant, $\gamma_j$ is the variant--exposure association, and $\Gamma_j$ is the variant--outcome association. Under valid instruments and no direct effect of the variants on the outcome, $\Gamma_j=\beta\gamma_j$. Thus the MR model used in the motivating literature is a special case of the summary-data IV framework above.

We measure average instrument strength by
$$\kappa=\frac{1}{L}\sum_{j=1}^L\frac{\gamma_j^2}{\sigma_{X_j}^2},$$
and estimate it by
$$\hat\kappa=\frac{1}{L}\sum_{j=1}^L\frac{\hat\gamma_j^2}{\hat\sigma_{X_j}^2}-1.$$

\begin{remark}[Scope of the generalisation]
All results below are stated for the summary-data IV model in Assumptions \ref{assume:structure}--\ref{assume:measure-error}. Consequently, the fixed-$L$ propositions and the many-instrument results do not require the instruments to be genetic variants. The many-instrument results do, however, rely on independence across instrument-specific summary estimates; they are not claims about arbitrary correlated-instrument two-sample IV.
\end{remark}

\section{Inverse-Variance Weighted (IVW) Estimator}

For each instrument $j$, the corresponding individual-instrument estimator is $\hat{\beta}_j = \hat{\Gamma}_j/\hat{\gamma}_j$. This is the summary data analogue of the individual-instrument estimator introduced in classical 2SLS theory. However, an estimator based on a single instrument may be unstable when $\hat{\gamma}_j$ is close to zero \citep{10.2307/2283473}. The inverse-variance weighted (IVW) estimator combines the $L$ individual-instrument estimators using inverse-variance weighting \citep{https://doi.org/10.1002/gepi.21758}:
$$\hat{\beta}_{\mathtt{IVW}} = \frac{\sum\limits_{j=1}^L\hat{w}_j\hat{\beta}_j}{\sum\limits_{j=1}^L\hat{w}_j},$$
where $\hat{w}_j = \hat{\gamma}_j^2/\hat{\sigma}_{Y_j}^2$ is the usual inverse-variance weight, obtained by treating $\hat{\gamma}_j$ as fixed. The IVW estimator can thus be written as
$$\hat{\beta}_{\mathtt{IVW}} = \frac{\sum\limits_{j=1}^L\hat{\Gamma}_j\hat{\gamma}_j/\hat{\sigma}_{Y_j}^2}{\sum\limits_{j=1}^L\hat{\gamma}_j^2/\hat{\sigma}_{Y_j}^2}.$$

Applying the classical 2SLS framework reveals the connection between IVW and 2SLS. For simplicity, consider individual-level data with exposure vector $\mathbf{x}$, outcome vector $\mathbf{y}$, and instrument matrix $\mathbf{Z}$. Let $\hat{\boldsymbol{\gamma}} = (\mathbf{Z}^T\mathbf{Z})^{-1}\mathbf{Z}^T\mathbf{x}$ and $\hat{\boldsymbol{\Gamma}} = (\mathbf{Z}^T\mathbf{Z})^{-1}\mathbf{Z}^T\mathbf{y}$.
The 2SLS criterion can be written as
\begin{align*}
\hat{\beta}_{\mathtt{2SLS}} &= \arg\min_{\beta}(\mathbf{y}-\mathbf{x}\beta)^T\mathbf{Z}(\mathbf{Z}^T\mathbf{Z})^{-1}\mathbf{Z}^T(\mathbf{y}-\mathbf{x}\beta) \\
&= \arg\min_{\beta}(\hat{\boldsymbol{\Gamma}}-\hat{\boldsymbol{\gamma}}\beta)^T(\mathbf{Z}^T\mathbf{Z})(\hat{\boldsymbol{\Gamma}}-\hat{\boldsymbol{\gamma}}\beta).
\end{align*}
Under homoscedasticity,
$$\widehat{\text{Var}}(\hat{\boldsymbol{\Gamma}}) = \hat{\sigma}_{V_Y}^2(\mathbf{Z}^T\mathbf{Z})^{-1}.$$
Hence, weighting the summary data associations by their inverse variances gives
$$\hat{\beta}_{\mathtt{IVW}} = \frac{\sum_{j=1}^L\hat{\Gamma}_j\hat{\gamma}_j/\hat{\sigma}_{Y_j}^2}{\sum_{j=1}^L\hat{\gamma}_j^2/\hat{\sigma}_{Y_j}^2}.$$
Thus, under the classical homoscedastic individual-level model, IVW corresponds to the 2SLS estimator expressed in terms of the instrument--treatment and instrument--outcome associations.

\subsection{Fixed-number-of-instruments asymptotics}\label{sec:ivw-fixedL}
The preceding connection motivates the form of the IVW estimator through classical 2SLS. However, the two-sample IVW estimator uses summary associations estimated in separate samples, so its asymptotic properties require separate analysis.

For the later variance comparison, we establish the fixed-$L$ consistency and asymptotic normality
of IVW directly under our summary-data model. Although fixed-instrument two-sample IV asymptotics
are well established \citep{10.1162/REST_a_00011,PACINI201650}, we provide explicit derivations in the present framework to obtain variance expressions directly comparable with dIVW and PS. Let $n=\min(n_X,n_Y)\to\infty$.

Since $L$ is fixed, the $\gamma_j$'s are not all zero by Assumption \ref{assume:structure}, and $\sigma_{X_j}^2 = \sigma_{\gamma_j}^2/n_X$, it follows that
$$\kappa = \frac{1}{L}\sum\limits_{j=1}^L\frac{\gamma_j^2}{\sigma_{X_j}^2} = \frac{1}{L}\sum\limits_{j=1}^L\frac{\gamma_j^2}{\sigma_{\gamma_j}^2/n_X} = \frac{n_X}{L}\sum\limits_{j=1}^L\frac{\gamma_j^2}{\sigma_{\gamma_j}^2} \to \infty.$$

\begin{proposition}\label{consis-Lfix} Suppose Assumptions \ref{assume:structure} and \ref{assume:measure-error} hold, and $L$ is fixed. Then $\hat{\beta}_{\mathtt{IVW}}$ is consistent, i.e. $\hat{\beta}_{\mathtt{IVW}} \overset{p}{\to} \beta$ as $n \to \infty$.
\end{proposition}

\begin{proposition}\label{asymnorm-Lfix} Suppose Assumptions \ref{assume:structure} and \ref{assume:measure-error} hold, and $L$ is fixed. Then $\hat{\beta}_{\mathtt{IVW}}$ is asymptotically normal, i.e. $$\frac{1}{\sqrt{V_{\mathtt{IVW}}}}(\hat{\beta}_{\mathtt{IVW}}-\beta) \overset{d}{\to} \mathcal{N}(0,1)$$ as $n \to \infty$, where
$$V_{\mathtt{IVW}} = \frac{V_1+V_2}{\left(\sum\limits_{j=1}^L\gamma_j^2/\sigma_{\Gamma_j}^2\right)^2}, ~V_1 = \frac{1}{an_Y}\sum\limits_{j=1}^L\frac{\gamma_j^2}{\sigma_{\Gamma_j}^2}\beta^2\frac{\sigma_{\gamma_j}^2}{\sigma_{\Gamma_j}^2} ~\mathrm{and}~ V_2 = \frac{1}{n_Y}\sum\limits_{j=1}^L\frac{\gamma_j^2}{\sigma_{\Gamma_j}^2}.$$
\end{proposition}

The proofs of both propositions are in Appendix \ref{app:proofs}.

Proposition \ref{asymnorm-Lfix} gives the asymptotic variance of the IVW estimator. We now derive the rates of its components.
\begin{align*}
    V_1 &= \frac{1}{an_Y}\sum\limits_{j=1}^L\frac{\gamma_j^2}{\sigma_{\Gamma_j}^2}\beta^2\frac{\sigma_{\gamma_j}^2}{\sigma_{\Gamma_j}^2} = O\bigg( \frac{1}{n_Y}\sum\limits_{j=1}^L \frac{\gamma_j^2}{n_Y\sigma_{Y_j}^2} \bigg) = O\Big( \frac{L\kappa}{n_Y^2} \Big) \\
    &= O\Big(\frac{1}{n}\Big) \qquad\text{since }L\text{ is a constant}, \\
    V_2 &= \frac{1}{n_Y}\sum\limits_{j=1}^L\frac{\gamma_j^2}{\sigma_{\Gamma_j}^2} = \Theta\Big(\frac{1}{n}\Big).
\end{align*}
Here, $V_1 = O(n^{-1})$ rather than $\Theta(n^{-1})$ to allow for $\beta=0$, in which case $V_1=0$. Since $V_1 \geq 0$ and $V_2 = \Theta(n^{-1})$, it follows that $V_1+V_2 = \Theta(n^{-1})$, and hence
$$\mathrm{Var}(\hat{\beta}_{\mathtt{IVW}}) = \frac{V_1+V_2}{\left(\sum\limits_{j=1}^L\gamma_j^2/\sigma_{\Gamma_j}^2\right)^2} = \Theta\Big(\frac{1}{n}\Big).$$
Hence, $\mathrm{Var}(\hat{\beta}_{\mathtt{IVW}}) \to 0$ as $n \to \infty$.

\subsection{Many weak instruments asymptotics}\label{IVW-manyL-theory}
To study the statistical properties of the IVW estimator when $L$ diverges ($L \to \infty$), we first rewrite the marginal estimates as the sum of true effects and errors, i.e.
$$\begin{pmatrix}
\hat{\Gamma}_j \\
\hat{\gamma}_j
\end{pmatrix} = \begin{pmatrix}
\Gamma_j \\
\gamma_j
\end{pmatrix}+\begin{pmatrix}
u_{Y_j} \\
u_{X_j}
\end{pmatrix}.$$

It follows from the assumption that
$$\begin{pmatrix}
u_{Y_j} \\
u_{X_j}
\end{pmatrix} \sim \mathcal{N}\left(\begin{pmatrix}
0 \\
0
\end{pmatrix},\begin{pmatrix}
\sigma_{Y_j}^2 & 0 \\
0 & \sigma_{X_j}^2
\end{pmatrix}\right).$$

\citet{10.1214/19-AOS1866} approximate the bias assuming $\sigma_{Y_j}^2 \equiv \sigma_{Y_1}^2$:
\[
  \begin{split}
    \hat{\beta}_{\mathtt{IVW}} &= \frac{\sum_{j=1}^L\hat{\Gamma}_j\hat{\gamma}_j}{\sum_{j=1}^L\hat{\gamma}_j^2} \approx \frac{\mathbb{E}[\sum_{j=1}^L\hat{\Gamma}_j\hat{\gamma}_j]}{\mathbb{E}[\sum_{j=1}^L\hat{\gamma}_j^2]} \\
    &= \frac{\beta\|\boldsymbol{\gamma}\|_2^2}{\|\boldsymbol{\gamma}\|_2^2+\sum_{j=1}^L\sigma_{X_j}^2} \approx \frac{\beta}{1+\kappa^{-1}}.
  \end{split}
\]
The last equality follows from $\mathbb{E}[\hat{\Gamma}_j\hat{\gamma}_j] = \mathbb{E}[\hat{\Gamma}_j]\mathbb{E}[\hat{\gamma}_j] = \Gamma_j\gamma_j = \beta\gamma_j^2$ and $\mathbb{E}[\hat{\gamma}_j^2] = \mathrm{Var}(\hat{\gamma}_j)+\mathbb{E}[\hat{\gamma}_j]^2 = \sigma_{X_j}^2+\gamma_j^2$. We defer the justification of the final approximation until the necessary machinery has been developed. Nevertheless, the approximate bias vanishes as $\kappa \to \infty$ (as is always the case when $L$ is fixed).

We now allow the number of instruments to diverge, $L \to \infty$. We write $\hat{\Gamma}_j = \Gamma_j+u_{Y_j} = \beta\gamma_j+u_{Y_j} = \beta(\hat{\gamma}_j-u_{X_j})+u_{Y_j} = \beta\hat{\gamma}_j+u_{Y_j}-\beta u_{X_j}$. For now, we follow \citet{10.1214/20-AOS2027} by assuming we know the variances of $\hat{\Gamma}_j$, so $\hat{\sigma}_{Y_j}^2 = \sigma_{Y_j}^2$.

The following results are due to \citet{10.1214/20-AOS2027}. We adapt and reorganise their presentation, drawing also on the exposition of \citet{10.1214/22-AOS2247}, and state the results without reproducing the proofs.

Before presenting the results, we introduce some notation. For two sequences of random variables $X$ and $Y$, we write $X \precsim_p Y$, or equivalently $X = O_p(Y)$ if the ratio $X/Y$ is bounded in probability. We write $X \prec_p Y$, or equivalently $X = o_p(Y)$ if $X/Y \overset{p}{\to} 0$. Finally, we write $X \asymp_p Y$, or equivalently $X = \Theta_p(Y)$ if both $X \precsim_p Y$ and $Y \precsim_p X$.

We now begin with the denominator $\sum_{j=1}^L\hat{\gamma}_j^2/\sigma_{Y_j}^2$.

\begin{lemma}\label{denom} Suppose Assumptions \ref{assume:structure} and \ref{assume:measure-error} hold, and $L \to \infty$. Then
$$\sum\limits_{j=1}^L\frac{\hat{\gamma}_j^2}{\sigma_{Y_j}^2} = \sum\limits_{j=1}^L\frac{\gamma_j^2+\sigma_{X_j}^2}{\sigma_{Y_j}^2}+O_p(\sqrt{L(\kappa+1)}) = (1+o_p(1))\sum\limits_{j=1}^L\frac{\gamma_j^2+\sigma_{X_j}^2}{\sigma_{Y_j}^2} \asymp_p L(\kappa+1).$$
\end{lemma}

\begin{remark} One might recall that when $L$ is fixed, we have $\sum_{j=1}^L\hat{\gamma}_j^2/n_Y\hat{\sigma}_{Y_j}^2 \overset{p}{\to} \sum_{j=1}^L\gamma_j^2/\sigma_{\Gamma_j}^2$. The two results are compatible because $\sum_{j=1}^L\sigma_{X_j}^2 = \sum_{j=1}^L\sigma_{\gamma_j}^2/n_X \to 0$ as $n \to \infty$ for any fixed $L$.
\end{remark}

\begin{assumption}\label{assume:iv-dist}
    The true instrument effects $\gamma_j$ satisfy
    $$\frac{\max_{1 \leq j \leq L}\gamma_j^2/\sigma_{X_j}^2}{\sum_{j=1}^L(\gamma_j^2/\sigma_{X_j}^2+1)} \to 0$$
    as $L \to \infty$.
\end{assumption}

Assumption \ref{assume:iv-dist} guarantees that no instrument is dominant. This condition is not required for the ordinary IVW result below under Assumption \ref{assume:measure-error}, but it will be required for dIVW. In particular, Assumption \ref{assume:measure-error} implies that $\sum_{j=1}^L\gamma_j(u_{Y_j}-\beta u_{X_j})/\sigma_{Y_j}^2$
is exactly normally distributed \citep{10.1214/22-AOS2247}.

\begin{lemma}\label{nom1} Suppose Assumptions \ref{assume:structure} and \ref{assume:measure-error} hold, and $L \to \infty$. Then
$$\frac{1}{\sqrt{\mathcal{V}_1}}\sum\limits_{j=1}^L\frac{\gamma_j(u_{Y_j}-\beta u_{X_j})}{\sigma_{Y_j}^2} \overset{d}{\to} \mathcal{N}(0,1),$$
where $\mathcal{V}_1 = \sum\limits_{j=1}^L\frac{\gamma_j^2}{\sigma_{Y_j}^2}\left(1+\beta^2\frac{\sigma_{X_j}^2}{\sigma_{Y_j}^2}\right) = \Theta(L\kappa)$. As a result,
$$\sum\limits_{j=1}^L\frac{\gamma_j(u_{Y_j}-\beta u_{X_j})}{\sigma_{Y_j}^2} = O_p(\sqrt{L\kappa}).$$
\end{lemma}

For the last term $\sum_{j=1}^Lu_{X_j}(u_{Y_j}-\beta u_{X_j})/\sigma_{Y_j}^2$, we have

\begin{lemma}\label{nom2} Suppose Assumptions \ref{assume:structure} and \ref{assume:measure-error} hold, and $L \to \infty$. Then
$$\frac{1}{\sqrt{\mathcal{V}_2}}\left(\sum\limits_{j=1}^L\frac{u_{X_j}(u_{Y_j}-\beta u_{X_j})}{\sigma_{Y_j}^2}+\beta\sum\limits_{j=1}^L\frac{\sigma_{X_j}^2}{\sigma_{Y_j}^2}\right) \overset{d}{\to} \mathcal{N}(0,1),$$
where $\mathcal{V}_2 = \sum\limits_{j=1}^L\frac{\sigma_{X_j}^2}{\sigma_{Y_j}^2}\left(1+2\beta^2\frac{\sigma_{X_j}^2}{\sigma_{Y_j}^2}\right) = \Theta(L)$. As a result,
$$\sum\limits_{j=1}^L\frac{u_{X_j}(u_{Y_j}-\beta u_{X_j})}{\sigma_{Y_j}^2} = -\beta\sum\limits_{j=1}^L\frac{\sigma_{X_j}^2}{\sigma_{Y_j}^2}+O_p(\sqrt{L}).$$
\end{lemma}

We can now state the main result for the IVW estimator, following Theorem S.1 in the supplementary material of \citet{10.1214/22-AOS2247}.

\begin{theorem}[IVW asymptotic properties]\label{asymnorm} Suppose Assumptions \ref{assume:structure} and \ref{assume:measure-error} hold, and $L \to \infty$. Then
$$\frac{1}{\sqrt{\mathsf{V}_{\mathtt{IVW}}}}\left[\hat{\beta}_{\mathtt{IVW}}-\beta\left(\frac{\sum_{j=1}^L\gamma_j^2/\sigma_{Y_j}^2}{\sum_{j=1}^L(\gamma_j^2+\sigma_{X_j}^2)/\sigma_{Y_j}^2}\right)\right] \overset{d}{\to} \mathcal{N}(0,1),$$
where $$\mathsf{V}_{\mathtt{IVW}} = \frac{\mathcal{V}_1+\mathcal{V}_2}{\left(\sum_{j=1}^L(\gamma_j^2+\sigma_{X_j}^2)/\sigma_{Y_j}^2\right)^2}.$$ As a result,
$$\hat{\beta}_{\mathtt{IVW}}-\beta \precsim_p \frac{1}{\kappa+1}+\sqrt{\frac{1}{L(\kappa+1)}}$$
and hence $\hat{\beta}_{\mathtt{IVW}}$ is consistent when $\kappa \to \infty$.
\end{theorem}

\begin{corollary}\label{asymnorm-cor} Suppose Assumptions \ref{assume:structure} and \ref{assume:measure-error} hold, and $L \to \infty$. If $\kappa/L \to \infty$, then $\hat{\beta}_{\mathtt{IVW}}$ is asymptotically normal, i.e.
$$\frac{1}{\sqrt{\mathsf{V}_{\mathtt{IVW}}}}(\hat{\beta}_{\mathtt{IVW}}-\beta) \overset{d}{\to} \mathcal{N}(0,1).$$
\end{corollary}

However, the assumption that $\kappa/L \to \infty$ is strong. \citet{10.1214/20-AOS2027} show that, approximately, $\kappa/L \leq n_X/L^2$. When the number of instruments is large relative to the first-stage sample size, the bound $\kappa/L \lesssim n_X/L^2$ illustrates how demanding the condition $\kappa/L \to \infty$ can be in practice. If $\kappa/L \precsim_p 1$, the asymptotic bias of the IVW estimator may be of first order or even dominating, invalidating the standard normal approximation.

\citet{10.1214/20-AOS2027} estimate $\mathsf{V}_{\mathtt{IVW}}$ by $\widehat{\mathsf{V}}_{\mathtt{IVW}}$
$$\widehat{\mathsf{V}}_{\mathtt{IVW}} = \frac{\sum_{j=1}^L\left[\hat{\gamma}_j^2/\hat{\sigma}_{Y_j}^2+\hat{\beta}_{\mathtt{IVW}}^2\hat{\sigma}_{X_j}^2/\hat{\sigma}_{Y_j}^2(\hat{\gamma}_j^2+\hat{\sigma}_{X_j}^2)/\hat{\sigma}_{Y_j}^2\right]}{\left(\sum_{j=1}^L\hat{\gamma}_j^2/\hat{\sigma}_{Y_j}^2\right)^2},$$
and establish consistency of this variance estimator.

\begin{theorem}[Consistent variance estimation]\label{convar} Suppose Assumptions \ref{assume:structure} and \ref{assume:measure-error} hold, and $L \to \infty$. If $\kappa/L \to \infty$, then 
$$\frac{\widehat{\mathsf{V}}_{\mathtt{IVW}}}{\mathsf{V}_{\mathtt{IVW}}} \overset{p}{\to} 1.$$
\end{theorem}

\section{Debiased IVW (dIVW) Estimator}

\subsection{The many weak instruments problem}

We now illustrate the finite-sample consequences of the many weak instruments problem for the IVW estimator.

\begin{example} Following the simulation design and performance measures of \citet{10.1214/22-AOS2247}, we generate 2,000 Monte Carlo samples with true causal effect $\beta = 0.2$, sample size $n:= n_X = n_Y = 100{,}000$, and $L = 200{,}000$ independent instruments. We assume homoskedastic variances $\sigma_{X_j}^2 = \sigma_{Y_j}^2 = 1/n$. We use the mixture distribution below to model sparse first-stage effects $\gamma_j$ (following the original MR simulation design):
$$\gamma \sim \frac{\pi}{2}\cdot\mathcal{TN}(0,\varepsilon_x^2;(-\infty,-a])+\frac{\pi}{2}\cdot\mathcal{TN}(0,\varepsilon_x^2;[a,+\infty))+(1-\pi)\cdot\delta_0,$$
where $\mathcal{TN}$ represents a truncated normal distribution with variance $\varepsilon_x^2 = 5 \times 10^{-5}$, and $\delta_0$ represents a Dirac measure centred on zero. The value $a$ is the truncation threshold that controls the instruments' strength. We set the mixing probability $\pi = 0.002$, so that 0.2\% of the instruments are truly correlated with the exposure. We illustrate the finite-sample performance of the two-sample IVW estimator in Table \ref{tab:tableIVW}.

\begin{table}[h]
\centering
\begin{tabular}{c|ccccc}
\hline
$a$ & $\hat{\beta}$ & Monte SD & SD & CP & Power \\
\hline
0 & 0.002 & 0.002 & 0.002 & 0 & 0.14 \\
0.005 & 0.004 & 0.002 & 0.002 & 0 & 0.39 \\
0.01 & 0.007 & 0.002 & 0.002 & 0 & 0.90 \\
0.015 & 0.012 & 0.002 & 0.002 & 0 & 1 \\
0.02 & 0.018 & 0.002 & 0.002 & 0 & 1 \\
\hline
\end{tabular}

\caption{IVW estimator under different truncation thresholds $a$.}

\textit{The true effect is $\beta = 0.2$. $\hat{\beta}$ denotes the IVW estimate; Monte SD and SD denote the empirical standard deviation and average estimated standard error, respectively. CP denotes the empirical coverage probability of the 95\% confidence interval, and Power denotes the proportion of $p$-values below 0.05.}
\label{tab:tableIVW}
\end{table}
\end{example}

The IVW estimates are severely attenuated toward zero, with greater attenuation for weaker instruments (smaller $a$). This reflects uncertainty in $\hat{\gamma}_j$. When its estimation error is non-negligible, Theorem \ref{asymnorm} shows that the asymptotic centre shifts from $\beta$ to
$$\beta\left(\frac{\sum_{j=1}^L\gamma_j^2/\sigma_{Y_j}^2}{\sum_{j=1}^L(\gamma_j^2+\sigma_{X_j}^2)/\sigma_{Y_j}^2}\right).$$

One approach to mitigate weak-instrument bias is to select only instruments that are sufficiently strongly associated with the exposure. However, selection based on estimated instrument strength can introduce additional problems, including the winner's curse \citep{10.1214/22-AOS2247}. We do not consider instrument selection further here. Instead, we next consider methods to address weak-instrument bias without such screening.

\subsection{Fixed-number-of-instruments asymptotics}

The debiased IVW (dIVW) estimator addresses the many weak instruments bias by correcting the denominator of the IVW estimator \citep{10.1214/20-AOS2027}:
\begin{align*}
\hat{\beta}_{\mathtt{dIVW}} &= \hat{\beta}_{\mathtt{IVW}}~\frac{\sum_{j=1}^L\hat{\gamma}_j^2/\hat{\sigma}_{Y_j}^2}{\sum_{j=1}^L(\hat{\gamma}_j^2-\hat{\sigma}_{X_j}^2)/\hat{\sigma}_{Y_j}^2}
\\ &= \frac{\sum_{j=1}^L\hat{\Gamma}_j\hat{\gamma}_j/\hat{\sigma}_{Y_j}^2}{\sum_{j=1}^L(\hat{\gamma}_j^2-\hat{\sigma}_{X_j}^2)/\hat{\sigma}_{Y_j}^2}.
\end{align*}
Thus, dIVW replaces $\hat{\gamma}_j^2$ in the IVW estimator by its bias-corrected counterpart $\hat{\gamma}_j^2-\hat{\sigma}_{X_j}^2$. We establish fixed-$L$ consistency and asymptotic normality for dIVW under the same summary-data framework. In particular, we show that the denominator bias correction does not affect the first-order asymptotic distribution, so IVW and dIVW have the same first-order asymptotic variance when $L$ is fixed.

\begin{proposition}\label{consis-Lfix-dIVW} Suppose Assumptions \ref{assume:structure} and \ref{assume:measure-error} hold, and $L$ is fixed. Then $\hat{\beta}_{\mathtt{dIVW}}$ is consistent, i.e. $\hat{\beta}_{\mathtt{dIVW}} \overset{p}{\to} \beta$ as $n \to \infty$.
\end{proposition}

\begin{proposition}\label{asymnorm-Lfix-dIVW} Suppose Assumptions \ref{assume:structure} and \ref{assume:measure-error} hold, and $L$ is fixed. Then $\hat{\beta}_{\mathtt{dIVW}}$ is asymptotically normal, i.e. $$\frac{1}{\sqrt{V_{\mathtt{dIVW}}}}(\hat{\beta}_{\mathtt{dIVW}}-\beta) \overset{d}{\to} \mathcal{N}(0,1)$$ as $n \to \infty$, where $V_{\mathtt{dIVW}} = V_{\mathtt{IVW}}$.
\end{proposition}

The proofs of both propositions are in Appendix \ref{app:proofs}.

\begin{remark} Although dIVW can exhibit greater finite-sample variability than IVW because the bias correction may increase denominator instability, the two estimators have the same first-order asymptotic variance when $L$ is fixed. The bias-correction term is of smaller order than the first-order stochastic variation.
\end{remark}

\subsection{Many weak instruments asymptotics}

For $L \to \infty$, we record the following results, which first appear in \citet{10.1214/20-AOS2027}.

\begin{theorem} Suppose Assumptions \ref{assume:structure}, \ref{assume:measure-error}, and \ref{assume:iv-dist} hold, and $L \to \infty$. If $\sqrt{L}\kappa \to \infty$, then $\hat{\beta}_{\mathtt{dIVW}}$ is asymptotically normal, i.e.
$$\frac{1}{\sqrt{\mathsf{V}_{\mathtt{dIVW}}}}(\hat{\beta}_{\mathtt{dIVW}}-\beta) \overset{d}{\to} \mathcal{N}(0,1),$$
where $$\mathsf{V}_{\mathtt{dIVW}} = \frac{\mathcal{V}_1+\mathcal{V}_2}{\left(\sum_{j=1}^L\gamma_j^2/\sigma_{Y_j}^2\right)^2}.$$

As a result,
$$\hat{\beta}_{\mathtt{dIVW}}-\beta \precsim_p \sqrt{\frac{\kappa+1}{L\kappa^2}}$$
and hence $\hat{\beta}_{\mathtt{dIVW}}$ is consistent.
\end{theorem}

We also have a consistent variance estimator.
\begin{theorem} Suppose Assumptions \ref{assume:structure}, \ref{assume:measure-error}, and \ref{assume:iv-dist} hold, and that $L \to \infty$. If $\sqrt{L}\kappa \to \infty$, then
$$\frac{\widehat{\mathsf{V}}_{\mathtt{dIVW}}}{\mathsf{V}_{\mathtt{dIVW}}} \overset{p}{\to} 1,$$
where the estimated variance is given by $$\widehat{\mathsf{V}}_{\mathtt{dIVW}} = \frac{\sum_{j=1}^L\left[\hat{\gamma}_j^2/\hat{\sigma}_{Y_j}^2+\hat{\beta}_{\mathtt{dIVW}}^2\hat{\sigma}_{X_j}^2/\hat{\sigma}_{Y_j}^2(\hat{\gamma}_j^2+\hat{\sigma}_{X_j}^2)/\hat{\sigma}_{Y_j}^2\right]}{\left(\sum_{j=1}^L(\hat{\gamma}_j^2-\hat{\sigma}_{X_j}^2)/\hat{\sigma}_{Y_j}^2\right)^2}.$$
\end{theorem}

The dIVW estimator requires a substantially weaker instrument strength condition than IVW: when $L \to \infty$, $\sqrt{L}\kappa \to \infty$ is substantially weaker than the condition $\kappa/L \to \infty$ for IVW, making dIVW more robust to the many weak instruments problem \citep{10.1214/20-AOS2027}.

In the many-instrument regime, this robustness comes with a bias-variance trade-off: dIVW corrects the weak-instrument bias of IVW at the cost of a larger asymptotic variance and potentially lower testing power \citep{10.1214/20-AOS2027}. In contrast, when $L$ is fixed, Proposition \ref{asymnorm-Lfix-dIVW} shows that IVW and dIVW have the same first-order asymptotic variance.

\section{Profile Score (PS) Estimator}
\subsection{Profile likelihood and score}
We introduce the profile likelihood/score approach proposed by \citet{10.1214/19-AOS1866}, which differs fundamentally from the IVW and dIVW approaches. The log-likelihood function of summary data (up to an additive constant) is
$$l(\beta,\gamma_1,\dots,\gamma_L) = -\frac{1}{2}\left[\sum\limits_{j=1}^L\frac{(\hat{\gamma}_j-\gamma_j)^2}{\sigma_{X_j}^2}+\sum\limits_{j=1}^L\frac{(\hat{\Gamma}_j-\gamma_j\beta)^2}{\sigma_{Y_j}^2}\right].$$
The only parameter of interest is $\beta$, so the other parameters $\boldsymbol{\gamma} := (\gamma_1,\dots,\gamma_L)$ are called nuisance parameters. We therefore treat $\boldsymbol{\gamma}$ as incidental nuisance parameters \citep{10.2307/1914288} and profile them out of the log-likelihood. \citet{10.1214/19-AOS1866} assume that the variances are known and that $\sigma_{Y_j}^2 \asymp 1/n_Y$ and $\sigma_{X_j}^2 \asymp 1/n_X$.

Profiling out $\boldsymbol{\gamma}$ yields the profile log-likelihood function of $\beta$:
$$l(\beta) = \max\limits_{\boldsymbol{\gamma}}l(\beta,\boldsymbol{\gamma}) = -\frac{1}{2}\sum\limits_{j=1}^L\frac{(\hat{\Gamma}_j-\beta\hat{\gamma}_j)^2}{\sigma_{X_j}^2\beta^2+\sigma_{Y_j}^2}.$$
The derivation of the profile log-likelihood is provided in Appendix \ref{app:proofs}.

Differentiating the profile log-likelihood gives the profile score
$$\psi(\beta) := l'(\beta) = \sum\limits_{j=1}^L\frac{(\hat{\Gamma}_j-\beta\hat{\gamma}_j)(\hat{\Gamma}_j\sigma_{X_j}^2\beta+\hat{\gamma}_j\sigma_{Y_j}^2)}{(\sigma_{X_j}^2\beta^2+\sigma_{Y_j}^2)^2}.$$

\subsection{Asymptotic properties}
Because the dimension of the nuisance parameter $\boldsymbol{\gamma}$ may increase with the number of instruments, standard fixed-dimensional likelihood arguments do not directly establish the large-sample behaviour of the PS estimator \citep{10.2307/1914288}. \citet{10.1214/19-AOS1866} show that PS nevertheless remains consistent and asymptotically normal under suitable strength and regularity conditions. They also show that the profile score need not satisfy the usual information identity, a form of information bias that affects likelihood-based inference \citep{MR1064420}.

For the profile-score asymptotics we impose bounded collective first-stage strength directly. In the MR formulation of \citet{10.1214/19-AOS1866}, this condition is motivated from bounded exposure variance and the distribution of genetic variants; for general two-sample IV it should be viewed as an explicit regularity condition on the sequence of first-stage coefficients.

\begin{assumption}[Bounded collective instrument strength]\label{assume:bounded-iv-strength}
    $\|\boldsymbol{\gamma}\|_2^2 = O(1)$.
\end{assumption}

Consequently, the average effect size converges to zero as $L \to \infty$,
$$\frac{1}{L}\sum\limits_{j=1}^L|\gamma_j| = \frac{1}{\sqrt{L}}\left(\frac{1}{\sqrt{L}}\|\boldsymbol{\gamma}\|_1\right) \leq \frac{1}{\sqrt{L}}\|\boldsymbol{\gamma}\|_2 \to 0.$$
Thus, Assumption \ref{assume:bounded-iv-strength} allows the individual first-stage coefficients $\gamma_j$ to be weak even when the instruments collectively contain sufficient first-stage information.

Let $n = \min(n_X,n_Y) \to \infty$.

\begin{theorem}[\citealp{10.1214/19-AOS1866}]\label{thm:consistency-simple} Suppose Assumptions \ref{assume:structure}, \ref{assume:measure-error}, and \ref{assume:bounded-iv-strength} hold, and $L/(n^2\|\boldsymbol{\gamma}\|_2^4)$
$\to 0$. Then the MLE $\hat{\beta}_{\mathtt{PS}}$ is consistent, i.e. $\hat{\beta}_{\mathtt{PS}} \overset{p}{\to} \beta$ as $n \to \infty$.
\end{theorem}

The condition involves $L/(n^2\|\boldsymbol{\gamma}\|_2^4) = \Theta(1/L\kappa^2)$, which we show and examine more closely in the next section. Importantly, the strength requirement is aggregate rather than instrument-specific. It therefore permits sparse first stages in which many individual coefficients $\gamma_j$ vanish, provided that the remaining instruments collectively supply sufficient information \citep{10.1214/19-AOS1866}.

The MLE $\hat{\beta}_{\mathtt{PS}}$ is a solution to the estimating equation $\psi(\hat{\beta}_{\mathtt{PS}}) = 0$. The Taylor expansion around $\beta$ gives us $0 = \psi(\hat{\beta}_{\mathtt{PS}}) = \psi(\beta)+\psi'(\beta)(\hat{\beta}_{\mathtt{PS}}-\beta)+\frac{1}{2}\psi''(\tilde{\beta})(\hat{\beta}_{\mathtt{PS}}-\beta)^2$, where $\tilde{\beta}$ lies between $\hat{\beta}_{\mathtt{PS}}$ and $\beta$. The following three lemmas are due to \citet{10.1214/19-AOS1866}.

\begin{lemma}\label{lem:1} If $\hat{\beta}$ is statistically consistent, then for a neighbourhood $\mathcal{N}$ of $\beta$,
$$\sup_{\beta_0 \in \mathcal{N}} (1/V_2) \psi''(\beta_0) = O_p(1).$$ In other words, $\frac{1}{2}\psi''(\tilde{\beta})(\hat{\beta}-\beta)^2$ is negligible.
\end{lemma}

\begin{lemma} \label{lem:2} Let $$V_1 = \sum_{j=1}^L \frac{\gamma_j^2 \sigma_{Y_j}^2 + \Gamma_j^2 \sigma_{X_j}^2 + \sigma_{X_j}^2 \sigma_{Y_j}^2}{(\sigma_{X_j}^2 \beta^2 + \sigma_{Y_j}^2)^2}.$$
Then $\psi(\beta) \overset{d}{\to} \mathcal{N}(0, V_1)$.
\end{lemma}

\begin{lemma} \label{lem:3} Let $$V_2 = V_1 - \sum\limits_{j=1}^L \frac{\sigma_{X_j}^2 \sigma_{Y_j}^2}{(\sigma_{X_j}^2 \beta^2 + \sigma_{Y_j}^2)^2} = \sum\limits_{j=1}^L \frac{\gamma_j^2 \sigma_{Y_j}^2 + \Gamma_j^2 \sigma_{X_j}^2}{(\sigma_{X_j}^2 \beta^2 + \sigma_{Y_j}^2)^2}.$$
Then $\psi'(\beta) \overset{p}{\to} -V_2$.
\end{lemma}

The profile score is called information unbiased if $V_1 = V_2$ \citep{MR1064420}. Combining the three lemmas and using Slutsky's theorem, we get

\begin{theorem}[\citealp{10.1214/19-AOS1866}] \label{thm:clt-simple}
  Suppose the assumptions in Theorem \ref{thm:consistency-simple} hold and at least one of the following conditions is satisfied as $n \to \infty$: (1) $L \to \infty$ and $\|\boldsymbol{\gamma}\|_3 / \|\boldsymbol{\gamma}\|_2 \to 0$; (2)
  $\kappa \to \infty$; then
  \begin{equation}
    \label{eq:clt-simple}
    \frac{V_2}{\sqrt{V_1}}(\hat{\beta}_{\mathtt{PS}} - \beta) \overset{d}{\to}
    \mathcal{N}(0, 1).
  \end{equation}
\end{theorem}

The theorem accommodates two qualitatively different sources of asymptotic information. With finitely many instruments, increasing sample size makes the instruments effectively strong through $\kappa\to\infty$; alternatively, when the number of instruments diverges, asymptotic normality can hold even as average instrument strength becomes small. In the latter regime, the norm condition limits concentration of the aggregate first-stage signal in a small number of instruments and provides the regularity needed for the central limit argument \citep{10.1214/19-AOS1866}.

To estimate $V_1$ and $V_2$ from the summary data, we must first estimate $\gamma_j^2$ and $\Gamma_j^2$. We replace $\gamma_j^2$ and $\Gamma_j^2$ with their unbiased sample estimates $\hat{\gamma}_j^2 - \sigma_{X_j}^2$ (seen in Lemma \ref{denom}) and $\hat{\Gamma}_j^2 - \sigma_{Y_j}^2$:

$$\hat{V}_1 = \sum_{j=1}^L \frac{(\hat{\gamma}_j^2 - \sigma_{X_j}^2) \sigma_{Y_j}^2 + (\hat{\Gamma}_j^2 - \sigma_{Y_j}^2) \sigma_{X_j}^2 + \sigma_{X_j}^2 \sigma_{Y_j}^2}{(\sigma_{X_j}^2 \hat{\beta}_{\mathtt{PS}}^2 + \sigma_{Y_j}^2)^2}$$

and $$\hat{V}_2 = \hat{V}_1 - \sum\limits_{j=1}^L \frac{\sigma_{X_j}^2 \sigma_{Y_j}^2}{(\sigma_{X_j}^2 \hat{\beta}_{\mathtt{PS}}^2 + \sigma_{Y_j}^2)^2} = \sum\limits_{j=1}^L \frac{(\hat{\gamma}_j^2 - \sigma_{X_j}^2) \sigma_{Y_j}^2 + (\hat{\Gamma}_j^2 - \sigma_{Y_j}^2) \sigma_{X_j}^2}{(\sigma_{X_j}^2 \hat{\beta}_{\mathtt{PS}}^2 + \sigma_{Y_j}^2)^2}.$$

\begin{theorem}[\citealp{10.1214/19-AOS1866}] \label{thm:consistent-variance-simple}
  Suppose the same assumptions in Theorem \ref{thm:clt-simple} hold. Then $\hat{V}_1 = V_1 (1 + o_p(1))$, $\hat{V}_2 = V_2 (1 + o_p(1))$, and
  \begin{equation}
    \label{eq:pivot-simple}
    \frac{\hat{V}_2}{\sqrt{\hat{V}_1}}(\hat{\beta}_{\mathtt{PS}} - \beta) \overset{d}{\to}
    \mathcal{N}(0, 1)~\mathrm{as}~n \to \infty.
  \end{equation}
\end{theorem}

Theorem \ref{thm:clt-simple} implies that $\mathrm{Var}(\hat{\beta}_{\mathtt{PS}}) \approx V_1/V_2^2$. Consider the rates of $V_1$ and $V_2$.
\[
  \begin{split}
    V_2 &= \sum\limits_{j=1}^L \frac{\gamma_j^2 \sigma_{Y_j}^2 + \Gamma_j^2 \sigma_{X_j}^2}
{(\sigma_{X_j}^2 \beta^2 + \sigma_{Y_j}^2)^2} = \Theta\bigg(\sum\limits_{j=1}^L \frac{\gamma_j^2 + \beta^2 \gamma_j^2}{(1/n)(\beta^2 + 1)^2}\bigg) = \Theta\bigg( n\sum\limits_{j=1}^L \frac{\gamma_j^2}{\beta^2 + 1} \bigg) \\
    &= \Theta(n\|\boldsymbol{\gamma}\|_2^2) = \Theta(L(n\|\boldsymbol{\gamma}\|_2^2/L)) = \Theta(L\kappa), \\
    V_1-V_2 &= \sum\limits_{j=1}^L \frac{\sigma_{X_j}^2 \sigma_{Y_j}^2}{(\sigma_{X_j}^2 \beta^2 + \sigma_{Y_j}^2)^2} = \Theta\bigg( \sum\limits_{j=1}^L \frac{1}{(\beta^2 + 1)^2} \bigg) = \Theta(L).
  \end{split}
\]
As $\kappa \to \infty$, the discrepancy between $V_1$ and $V_2$ becomes asymptotically negligible relative to $V_2$. More generally, the relative contribution of information bias is of order $\kappa^{-1}$ \citep{10.1214/19-AOS1866}. Thus, $$\mathrm{Var}(\hat{\beta}_{\mathtt{PS}}) \approx \frac{V_1}{V_2^2} = \frac{V_2+\Theta(L)}{V_2}\frac{1}{V_2} = \frac{1+\Theta(L)/V_2}{V_2} = \frac{1+\Theta(L)/\Theta(L\kappa)}{V_2} = \frac{1+\Theta(\kappa^{-1})}{V_2}.$$
To illustrate, we look at the case of $\beta = 0$ and $\sigma_{Y_j}^2 \equiv \sigma_{Y_1}^2$. Since $\beta = 0$, $\Gamma_j = \beta\gamma_j = 0$ for all $j \in [L]$. Then
\[
  \begin{split}
    V_2 &= \sum\limits_{j=1}^L \frac{\gamma_j^2 \sigma_{Y_j}^2}{(\sigma_{Y_j}^2)^2} = \sum\limits_{j=1}^L \frac{\gamma_j^2}{\sigma_{Y_j}^2} = \frac{1}{\sigma_{Y_1}^2}\sum\limits_{j=1}^L \gamma_j^2~\mathrm{and} \\
    V_1 &= V_2+\sum\limits_{j=1}^L \frac{\sigma_{X_j}^2 \sigma_{Y_j}^2}{(\sigma_{Y_j}^2)^2} = V_2+\sum\limits_{j=1}^L \frac{\sigma_{X_j}^2}{\sigma_{Y_j}^2} = V_2+\frac{1}{\sigma_{Y_1}^2}\sum\limits_{j=1}^L \sigma_{X_j}^2.
  \end{split}
\]
Thus, $V_1/V_2 = 1+(\sum_{j=1}^L \sigma_{X_j}^2)/(\sum_{j=1}^L \gamma_j^2)$. If, for example, we impose $\sigma_{X_j}^2 \equiv \sigma_{X_1}^2$, then the second term becomes $L/(\sum_{j=1}^L\gamma_j^2/\sigma_{X_1}^2) = \kappa^{-1}$. Hence $V_1/V_2^2 = (1+\kappa^{-1})/V_2$. If instead we assume that $\sigma_{Y_j}^2/\sigma_{X_j}^2$ is independent of $j$, we can also obtain $\mathrm{Var}(\hat{\beta}_{\mathtt{PS}}) \propto \frac{1 + \kappa^{-1}}{L \kappa}$. We can use this approximation to choose the optimal number of instruments $L$ \citep{10.1214/19-AOS1866}.

In the MR literature, \citet{10.1214/19-AOS1866} extend the PS framework to violations of the valid-IV restriction through adjusted and robust adjusted profile scores. In general IV terminology, such extensions address particular forms of instrument invalidity or direct effects. Our comparison concerns the valid-IV proportionality model in Assumption \ref{assume:structure}, so we restrict attention to PS.

\section{Comparison of Asymptotic Conditions}\label{sec-rates}
In this section, we compare and analyse the conditions under which the estimators are consistent and asymptotically normal. The theoretical properties of the IVW estimator under the many weak instruments regime were not fully understood until \citet{10.1214/20-AOS2027}, who characterise its behaviour according to the asymptotic average instrument strength as $L \to \infty$. The main implications of their results are summarised below, where ``not guaranteed'' means that the stated condition alone is insufficient to establish asymptotic normality.
\begin{center}
\begin{tabular}{||c c | c c||} 
 \hline
 Case & Assumptions & Consistency & Asymptotic Normality \\ [0.5ex] 
 \hline\hline
 1 & $\kappa/L \to \infty$ & Yes & Yes \\
 \hline
 2 & $\kappa \to \infty$ alone & Yes & Not guaranteed \\
 \hline
 3 & $\kappa \to c > 0$ & Not in general & No \\
 \hline
 4 & $\kappa \to 0$ & Not in general & No \\[1ex] 
 \hline
\end{tabular}
\end{center}

The inconsistency statements for bounded average instrument strength concern the general case $\beta\neq 0$. When $\beta=0$, the attenuation bias vanishes, and IVW can remain consistent even if $\kappa$ does not diverge. However, this does not by itself establish the usual asymptotic normal approximation.

To compare the estimators in the many-instrument regime, we list below sufficient conditions for consistency and asymptotic normality when $L \to \infty$. Conditions (i)--(vii) concern, respectively, the sample sizes, number of instruments, variance estimation, variance behaviour, collective instrument strength, instrument concentration, and additional conditions specific to PS. We do not list Assumptions \ref{assume:structure} and \ref{assume:measure-error} below because all three estimators require them.
\begin{center}
\renewcommand{\arraystretch}{1.2}
\begin{tabular}{|c|ccc|} 
 \hline
 Conditions & IVW & dIVW & PS \\ [0.5ex] 
 \hline
 (i) & $n_X/n_Y \to a$ & $n_X/n_Y \to a$ & $n = \min(n_X,n_Y)$ \\ 
  & $n \to \infty$ & $n \to \infty$ & $n \to \infty$ \\
 \hline
 (ii) & $L \to \infty$ & $L \to \infty$ &  \\
 \hline
 (iii) & $\sigma_{X_j}^2, \sigma_{Y_j}^2$ known & $\sigma_{X_j}^2, \sigma_{Y_j}^2$ known & $\sigma_{X_j}^2, \sigma_{Y_j}^2$ known \\
  & & or $L/n_X \to 0$ &  \\
 \hline
 (iv) & $0 < \sigma_{X_j}^2/\sigma_{Y_j}^2 < \infty$ & $0 < \sigma_{X_j}^2/\sigma_{Y_j}^2 < \infty$ & $\sigma_{X_j}^2, \sigma_{Y_j}^2 = \Theta(1/n)$ \\
 \hline
 (v) & $\kappa/L \to \infty$ & $\sqrt{L}\kappa \to \infty$ & $L/(n^2\|\boldsymbol{\gamma}\|_2^4) \to 0$ \\
 \hline
 (vi) &  & $\dfrac{\max_j\gamma_j^2/\sigma_{X_j}^2}{(\kappa+1)L} \to 0$ &  \\
 \hline
 (vii) &  &  & $\|\boldsymbol{\gamma}\|_2^2 = O(1)$, and \\
 & & & either $L \to \infty$ with \\
 & & & $\|\boldsymbol{\gamma}\|_3/\|\boldsymbol{\gamma}\|_2 \to 0$, \\
  & & & or $\kappa \to \infty$ \\[1ex] 
 \hline
\end{tabular}
\end{center}

As discussed in Section \ref{IVW-manyL-theory}, IVW does not require the no-dominating-instrument condition under Assumption \ref{assume:measure-error}. In contrast, dIVW requires condition (vi)
$$\frac{\max_j \gamma_j^2/\sigma_{X_j}^2}{(\kappa+1)L}\to 0.$$

We first observe that the collective instrument strength conditions in (v) for dIVW and PS are equivalent under the variance assumptions of Zhao et al.

\begin{proposition}[Equivalence of collective-strength conditions]
\label{prop:strength-equivalence}
Suppose that $\sigma_{X_j}^2=\Theta(n^{-1})$
uniformly over $j=1,\dots,L$. Then
\[
\frac{L}{n^2\|\boldsymbol{\gamma}\|_2^4}
\to 0
\quad\Longleftrightarrow\quad
\sqrt{L}\kappa\to\infty.
\]
\end{proposition}

\begin{proof}
Since
$$\kappa = \frac{1}{L}\sum\limits_{j=1}^L\frac{\gamma_j^2}{\sigma_{X_j}^2} = \frac{1}{L}\sum\limits_{j=1}^L\frac{\gamma_j^2}{\Theta(1/n)} = \frac{1}{L}\Theta\left(n\sum\limits_{j=1}^L\gamma_j^2\right) = \Theta\left(\frac{n\|\boldsymbol{\gamma}\|_2^2}{L}\right),$$ it follows that
$$\frac{L}{n^2\|\boldsymbol{\gamma}\|_2^4} = \frac{L}{(n\|\boldsymbol{\gamma}\|_2^2)^2} = \frac{L}{L^2(n\|\boldsymbol{\gamma}\|_2^2/L)^2} = \frac{1}{L\Theta(\kappa)^2} = \Theta\left(\frac{1}{L\kappa^2}\right)$$ as stated by \citet{10.1214/19-AOS1866}. Hence
$$\frac{L}{n^2\|\boldsymbol{\gamma}\|_2^4} \to 0 \quad \Longleftrightarrow \quad \sqrt{L}\kappa \to \infty.$$
\end{proof}

Thus, the condition on collective instrument strength in (v) required for consistency of PS coincides with the corresponding condition used to establish consistency and asymptotic normality of dIVW. In contrast, the sufficient condition given by \citet{10.1214/20-AOS2027} for consistency and asymptotic normality of IVW is $\kappa/L \to \infty$, which is considerably stronger. We therefore focus henceforth on comparing dIVW and PS.

To gain insights into the remaining conditions, particularly (vi) and (vii), we consider a sparse setting with one relatively strong instrument and a growing number of weaker instruments. Assume that $n \to \infty$, $\sigma_{X_j}^2, \sigma_{Y_j}^2$ are known, and their ratios are uniformly bounded above and away from zero. We parametrise the strength of the relatively strong instrument by $k < 1/2$, which ensures the common collective-strength condition $\sqrt{L}\kappa \to \infty$ in the example below. For brevity, throughout the example below, we say that an estimator is ``valid'' when the relevant sufficient conditions for its consistency and asymptotic normality are satisfied.

\begin{example}[one strong, some weak, all other zero]\label{ex:rate-compare-1strong-someweak}
    Suppose $\gamma_1, \gamma_2, \dots, \gamma_{r+1} \neq 0$ and $\gamma_{r+2} = \cdots = \gamma_L = 0$, where $1 \leq r \leq L-1$. We can write $r = L^m$, $0 \leq m < 1$. Consider $L \to \infty$, i.e. $L = \Theta(n^l)$ for some $l > 0$. We can take $\gamma_1^2/\sigma_{X_1}^2 \asymp L/L^k$ and $\gamma_2^2/\sigma_{X_2}^2, \dots, \gamma_{r+1}^2/\sigma_{X_{r+1}}^2 \asymp L/L^{1/2}$. Then    
    $$\kappa \asymp \frac{1}{L}\left(\frac{L}{L^k}+r\frac{L}{L^{1/2}}\right) = \frac{1}{L^k}+\frac{r}{L^{1/2}}=L^{-k}+L^{m-1/2}.$$
    Hence, $\sqrt{L}\kappa \asymp L^{1/2-k}+L^m \to \infty$.
    
    For dIVW, the no-dominating-instrument condition becomes
    $$\frac{\max_j\gamma_j^2/\sigma_{X_j}^2}{(\kappa+1)L} \asymp \frac{L/L^k}{(L^{-k}+L^{m-1/2}+1)L} = \frac{1}{1+L^{m-1/2+k}+L^k}.$$
    Thus, the condition holds if either $m > 1/2-k$ or $k > 0$. Equivalently, if $k < 0$, we require $m > 1/2-k$, which is feasible only for $-1/2 < k < 0$; if $k = 0$, we require $m > 1/2$; if $0 < k < 1/2$, the condition holds automatically.
    
    For PS,
    $$\|\gamma\|_2^2 = \gamma_1^2+\sum_{j=2}^{r+1}\gamma_j^2 \asymp \frac{L^{1-k}+rL^{1/2}}{n} \asymp \frac{n^{(1-k)l}}{n}+\frac{n^{(m+1/2)l}}{n}.$$
    For this to be $O(1)$, we require
    $$l \leq \min\left(\frac{1}{1-k},\frac{1}{1/2+m}\right).$$ 
    Now,
    $$\|\gamma\|_3^3 = |\gamma_1|^3+\sum_{j=2}^{r+1}|\gamma_j|^3 \asymp \frac{L^{3(1-k)/2}+rL^{3/4}}{n^{3/2}}.$$
Then,
$$\frac{\|\boldsymbol{\gamma}\|_3}{\|\boldsymbol{\gamma}\|_2} \asymp
\frac{\left(L^{3(1-k)/2}+L^{m+3/4}\right)^{1/3}}{\left(L^{1-k}+L^{m+1/2}\right)^{1/2}}.$$
For this ratio to converge to zero, the contribution
of the relatively strong instrument must be negligible.
This requires
$$1-k<m+\frac{1}{2},$$
or equivalently $m>1/2-k$. Under this condition, the denominator is of order
$L^{m/2+1/4}$. The contribution of the weaker
instruments to the ratio is therefore of order
$$\frac{L^{m/3+1/4}}{L^{m/2+1/4}} = L^{-m/6}\to0,$$
since $m>1/2-k>0$ for $k<1/2$. Hence,
$$\frac{\|\boldsymbol{\gamma}\|_3}{\|\boldsymbol{\gamma}\|_2}\to0
\quad\Longleftrightarrow\quad
m>\frac12-k.$$
    Also, $\kappa = 1/L^k+L^{m-1/2} \to \infty$ if $k < 0$ or $m > 1/2$. Thus, one of the additional PS normality conditions is satisfied if $m > 1/2-k$, $k < 0$, or $m > 1/2$. Together with the bounded-strength condition, PS therefore additionally requires $l \leq \min\left(\dfrac{1}{1-k},\dfrac{1}{1/2+m}\right)$.
    
    The comparison is more nuanced in this case, so we summarise the validity conditions for dIVW and PS across different ranges of $k$:

\begin{table}[!h]
\centering
\begin{tabular}{c|cc}
\hline
Range of $k$ & dIVW & PS \\
\hline
$k<0$ 
& $-\frac{1}{2}<k<0$ and $m>\frac{1}{2}-k$ 
& $l \leq \min\left(\dfrac{1}{1-k},\dfrac{1}{1/2+m}\right)$ \\[2pt]
$k=0$ 
& $m>\frac{1}{2}$ 
& $m>\frac{1}{2}$ and $l \leq \dfrac{1}{1/2+m}$ \\[2pt]
$0<k<\frac{1}{2}$ 
& Valid 
& $m>\frac{1}{2}-k$ and $l \leq \min\left(\dfrac{1}{1-k},\dfrac{1}{1/2+m}\right)$ \\
\hline
\end{tabular}
\caption{Additional validity conditions for dIVW and PS across different ranges of $k$.}
\label{tab:validity_conditions}
\end{table}

Thus, neither set of sufficient conditions nests the other. When $k<0$, PS can satisfy its normality condition through $\kappa\to\infty$, even when the single stronger instrument prevents the dIVW no-dominating-instrument condition from holding, although PS remains subject to its restriction on the growth of $L$ relative to $n$. Conversely, when $0<k<1/2$, the dIVW no-dominating-instrument condition holds automatically, whereas PS requires sufficiently many weak non-zero instruments, $m>1/2-k$, together with its growth-rate restriction. Increasing the number of weak non-zero instruments can therefore expand the parameter region in which the PS concentration condition is satisfied.
\end{example}

The sufficient conditions therefore differ beyond their common collective-strength requirement. In particular, dIVW imposes a no-dominating-instrument condition expressed in terms of the largest standardised first-stage coefficient, whereas the many-instrument PS normality result controls concentration through $\|\boldsymbol{\gamma}\|_3/\|\boldsymbol{\gamma}\|_2\to0$. Moreover, conditional on its other regularity conditions, dIVW does not impose a separate upper growth-rate restriction on $L$ relative to $n$. By contrast, PS combines $L/(n^2\|\boldsymbol{\gamma}\|_2^4)\to0$ with $\|\boldsymbol{\gamma}\|_2^2=O(1)$, which in particular implies $L=o(n^2)$. Thus, although the principal collective-strength conditions for dIVW and PS can be expressed equivalently under the PS variance scaling, their remaining regularity conditions are not directly nested.

\section{Comparison of Asymptotic Variances}\label{sec-var-compare}
We now compare the variance formulae for IVW/dIVW and PS under both fixed and diverging numbers of instruments. We first consider the fixed-$L$ setting, where IVW and dIVW have the same first-order asymptotic variance, and then extend the comparison to $L\to\infty$, where their variance formulae differ. \citet{10.1214/19-AOS1866} show that the profile score is generally information biased: the variance of the profile score and the negative expected derivative of the score do not coincide. The resulting additional variance component is analogous to variance adjustments arising under many weak moment conditions for continuously updated GMM (CUE), the GMM analogue of LIML in the linear IV setting \citep{Hansen01071996,https://doi.org/10.3982/ECTA6224}. Our aim here is to compare the three variance formulae across the two asymptotic regimes, identifying their relative orders and, under a common variance-ratio condition, their exact relationships. These comparisons concern asymptotic variance formulae rather than higher-order expansions of the actual sampling variances or an unconditional ranking of estimator efficiency.

\subsection{Fixed number of instruments}
\subsubsection{Variance decomposition and asymptotic orders}
We begin with the fixed-$L$ setting, in which the first-order asymptotic variances of IVW and dIVW coincide. We compare this common variance with the PS variance formula and examine the order of the additional component arising from profile-score information bias. To establish this comparison, we first recall the variance expressions established in Proposition \ref{asymnorm-Lfix}, Lemmas \ref{lem:2} and \ref{lem:3}, and Theorem \ref{thm:clt-simple}.
$$V_{\mathtt{IVW}} = \frac{V_1+V_2}{\left(\sum\limits_{j=1}^L\gamma_j^2/\sigma_{\Gamma_j}^2\right)^2}, ~V_1 = \frac{1}{an_Y}\sum\limits_{j=1}^L\frac{\gamma_j^2}{\sigma_{\Gamma_j}^2}\beta^2\frac{\sigma_{\gamma_j}^2}{\sigma_{\Gamma_j}^2} ~\mathrm{and}~ V_2 = \frac{1}{n_Y}\sum\limits_{j=1}^L\frac{\gamma_j^2}{\sigma_{\Gamma_j}^2}.$$
$$\tilde{V}_1 = \sum_{j=1}^L \frac{\gamma_j^2 \sigma_{Y_j}^2 + \Gamma_j^2 \sigma_{X_j}^2 + \sigma_{X_j}^2 \sigma_{Y_j}^2}{(\sigma_{X_j}^2 \beta^2 + \sigma_{Y_j}^2)^2}.$$
$$\tilde{V}_2 = \tilde{V}_1 - \sum\limits_{j=1}^L \frac{\sigma_{X_j}^2 \sigma_{Y_j}^2}{(\sigma_{X_j}^2 \beta^2 + \sigma_{Y_j}^2)^2} = \sum\limits_{j=1}^L \frac{\gamma_j^2 \sigma_{Y_j}^2 + \Gamma_j^2 \sigma_{X_j}^2}{(\sigma_{X_j}^2 \beta^2 + \sigma_{Y_j}^2)^2}.$$
$$V_{\mathtt{PS}} = \frac{\tilde{V}_1}{\tilde{V}_2^2}.$$
To facilitate the comparison with $V_{\mathtt{IVW}}$, we decompose $V_{\mathtt{PS}}$ into two components. This decomposition is an algebraic rewriting of the information-bias relation $\tilde V_1-\tilde V_2>0$ in the profile-score theory, rather than a claim that the existence of an additional PS variance component is new.
\begin{lemma}[PS variance decomposition]\label{lem:PS-var-decomp}
The variance of the profile score estimator admits the decomposition
$$V_{\mathtt{PS}} = V_{\mathtt{PS}}^{(1)} + V_{\mathtt{PS}}^{(2)},$$
where
\begin{align*}
V_{\mathtt{PS}}^{(1)} &:= \frac{1}{\tilde V_2}
=\left(\sum_{j=1}^L\frac{\gamma_j^2}{\sigma_{X_j}^2\beta^2+\sigma_{Y_j}^2}\right)^{-1},\\
V_{\mathtt{PS}}^{(2)} &:= \frac{1}{\tilde V_2^2}\sum_{j=1}^L
\frac{\sigma_{X_j}^2\sigma_{Y_j}^2}{(\sigma_{X_j}^2\beta^2+\sigma_{Y_j}^2)^2}\geq0.
\end{align*}
\end{lemma}

\begin{proof}
Since $\Gamma_j=\beta\gamma_j$, we have
\begin{align*}
    \tilde{V}_2 &= \sum_{j=1}^L \frac{\gamma_j^2 \sigma_{Y_j}^2 + \Gamma_j^2 \sigma_{X_j}^2}{(\sigma_{X_j}^2 \beta^2 + \sigma_{Y_j}^2)^2} \\
    &= \sum_{j=1}^L \frac{\gamma_j^2 \sigma_{Y_j}^2 + \beta^2\gamma_j^2 \sigma_{X_j}^2}{(\sigma_{X_j}^2 \beta^2 + \sigma_{Y_j}^2)^2} \\
    &= \sum_{j=1}^L \frac{\gamma_j^2(\sigma_{Y_j}^2 + \beta^2 \sigma_{X_j}^2)}{(\sigma_{X_j}^2 \beta^2 + \sigma_{Y_j}^2)^2} \\
    &= \sum_{j=1}^L \frac{\gamma_j^2}{\sigma_{X_j}^2 \beta^2 + \sigma_{Y_j}^2}.
\end{align*}
Then,
\begin{align*}
    \tilde{V}_1 &= \tilde{V}_2 + \sum\limits_{j=1}^L \frac{\sigma_{X_j}^2 \sigma_{Y_j}^2}{(\sigma_{X_j}^2 \beta^2 + \sigma_{Y_j}^2)^2} \\
    &= \sum_{j=1}^L \left[\frac{\gamma_j^2}{\sigma_{X_j}^2 \beta^2 + \sigma_{Y_j}^2} + \frac{\sigma_{X_j}^2 \sigma_{Y_j}^2}{(\sigma_{X_j}^2 \beta^2 + \sigma_{Y_j}^2)^2}\right].
\end{align*}
Therefore,
\begin{align*}
    V_{\mathtt{PS}} &= \frac{\tilde{V}_1}{\tilde{V}_2^2} \\
    &= \frac{1}{\tilde{V}_2^2}\left[\tilde{V}_2 + \sum\limits_{j=1}^L \frac{\sigma_{X_j}^2 \sigma_{Y_j}^2}{(\sigma_{X_j}^2 \beta^2 + \sigma_{Y_j}^2)^2}\right] \\
    &= \frac{1}{\tilde{V}_2} + \frac{1}{\tilde{V}_2^2}\sum\limits_{j=1}^L \frac{\sigma_{X_j}^2 \sigma_{Y_j}^2}{(\sigma_{X_j}^2 \beta^2 + \sigma_{Y_j}^2)^2} \\
    &= V_{\mathtt{PS}}^{(1)} + V_{\mathtt{PS}}^{(2)}.
\end{align*}
\end{proof}

\begin{proposition}[Fixed-$L$ variance orders]\label{prop:fixed-L-variance-orders}
Suppose $L$ is fixed, $n_X/n_Y\to a\in(0,\infty)$, and the fixed-$L$ variance scaling in the summary-data model holds. Then
$$V_{\mathtt{PS}}^{(1)}=\Theta(n^{-1}),\qquad
V_{\mathtt{PS}}^{(2)}=\Theta(n^{-2}),\qquad
V_{\mathtt{IVW}}=\Theta(n^{-1}).$$
Consequently,
$$\frac{V_{\mathtt{PS}}^{(2)}}{V_{\mathtt{IVW}}}=\Theta(n^{-1})\to0.$$
\end{proposition}

The proof is in Appendix \ref{app:proofs}.

Thus, in the fixed-$L$ setting, the additional PS variance component is of smaller order than both the leading PS component and the IVW variance.

\subsubsection{Exact variance ordering and asymptotic equivalence}
We next impose additional structure under which the IVW and PS variances can be compared exactly. Since $V_{\mathtt{IVW}}$ is expressed in terms of $\sigma_{\Gamma_j}^2$ and $\sigma_{\gamma_j}^2$, whereas $V_{\mathtt{PS}}$ is expressed in terms of $\sigma_{Y_j}^2$ and $\sigma_{X_j}^2$, we first re-express $V_{\mathtt{IVW}}$ in terms of the latter quantities. Recall that the true variances are defined by
$$\sigma_{Y_j}^2:=\frac{\sigma_{\Gamma_j}^2}{n_Y},
\qquad
\sigma_{X_j}^2:=\frac{\sigma_{\gamma_j}^2}{n_X},
\qquad j=1,\dots,L.$$
For the exact variance comparison, we strengthen the asymptotic condition $n_X/n_Y\to a$ to the finite-sample condition $n_X/n_Y=a$, equivalently $n_X=an_Y$. Hence,
\begin{align*}
    V_1 &= \frac{1}{n_X}\sum\limits_{j=1}^L\frac{\gamma_j^2}{n_Y\sigma_{Y_j}^2}\beta^2\frac{n_X\sigma_{X_j}^2}{n_Y\sigma_{Y_j}^2} = \frac{1}{n_Y^2}\sum\limits_{j=1}^L\frac{\gamma_j^2}{\sigma_{Y_j}^2}\beta^2\frac{\sigma_{X_j}^2}{\sigma_{Y_j}^2} \\
    V_2 &= \frac{1}{n_Y}\sum\limits_{j=1}^L\frac{\gamma_j^2}{n_Y\sigma_{Y_j}^2} = \frac{1}{n_Y^2}\sum\limits_{j=1}^L\frac{\gamma_j^2}{\sigma_{Y_j}^2} \\
    V_1+V_2 &= \frac{1}{n_Y^2}\sum\limits_{j=1}^L\frac{\gamma_j^2}{\sigma_{Y_j}^2}\left(\beta^2\frac{\sigma_{X_j}^2}{\sigma_{Y_j}^2}+1\right) = \frac{1}{n_Y^2}\sum\limits_{j=1}^L\frac{\gamma_j^2}{\sigma_{Y_j}^2}\frac{\beta^2\sigma_{X_j}^2+\sigma_{Y_j}^2}{\sigma_{Y_j}^2} \\
    \left(\sum\limits_{j=1}^L\frac{\gamma_j^2}{\sigma_{\Gamma_j}^2}\right)^2 &= \left(\sum\limits_{j=1}^L\frac{\gamma_j^2}{n_Y\sigma_{Y_j}^2}\right)^2 = \frac{1}{n_Y^2}\left(\sum\limits_{j=1}^L\frac{\gamma_j^2}{\sigma_{Y_j}^2}\right)^2
\end{align*}
Therefore, $$V_{\mathtt{IVW}} = \dfrac
{\displaystyle\sum\limits_{j=1}^L\dfrac{\gamma_j^2}{\sigma_{Y_j}^2}\dfrac{\beta^2\sigma_{X_j}^2+\sigma_{Y_j}^2}{\sigma_{Y_j}^2}}
{\displaystyle\left(\sum\limits_{j=1}^L\dfrac{\gamma_j^2}{\sigma_{Y_j}^2}\right)^2}.$$

We next derive the exact variance expressions under a common variance ratio condition. These expressions show, in particular, that the leading PS variance component coincides exactly with the IVW variance, while the second PS variance component is strictly positive.
\begin{lemma}\label{lem:exact-variance-expressions}
    Suppose $n_X/n_Y = a$ and, for some constant $c > 0$,
    $$\sigma_{X_j}^2 = c\sigma_{Y_j}^2, \qquad j = 1,\dots,L.$$
    Then, for every $L \geq 1$,
    $$V_{\mathtt{PS}}^{(1)} = V_{\mathtt{IVW}} = \frac{c\beta^2+1}{\sum_{j=1}^L\gamma_j^2/\sigma_{Y_j}^2},$$
    and $$V_{\mathtt{PS}}^{(2)} = \frac{cL}{\left(\sum_{j=1}^L\gamma_j^2/\sigma_{Y_j}^2\right)^2} > 0.$$
\end{lemma}

The proof is provided in Appendix \ref{app:proofs}.

\begin{proposition}[Fixed-$L$ variance formula comparison]\label{prop:exact-var-order}
Under the conditions of Lemma \ref{lem:exact-variance-expressions},
$$V_{\mathtt{PS}}=V_{\mathtt{IVW}}+V_{\mathtt{PS}}^{(2)}>V_{\mathtt{IVW}}.$$
Since $V_{\mathtt{dIVW}}=V_{\mathtt{IVW}}$ to first order for fixed $L$, the same comparison holds between the PS asymptotic variance and the common first-order IVW/dIVW variance.
\end{proposition}

\begin{proof}
By Lemma \ref{lem:exact-variance-expressions}, $$V_{\mathtt{PS}}^{(1)} = V_{\mathtt{IVW}} \qquad\text{and}\qquad V_{\mathtt{PS}}^{(2)} > 0.$$
Together with the decomposition in Lemma \ref{lem:PS-var-decomp}, this gives
$$V_{\mathtt{PS}} = V_{\mathtt{PS}}^{(1)}+V_{\mathtt{PS}}^{(2)} = V_{\mathtt{IVW}}+V_{\mathtt{PS}}^{(2)} > V_{\mathtt{IVW}}.$$
\end{proof}

\begin{remark}
    Since $n_Y\sigma_{Y_j}^2 = \sigma_{\Gamma_j}^2$ and $n_X\sigma_{X_j}^2 = \sigma_{\gamma_j}^2$, the condition $\sigma_{X_j}^2 = c\sigma_{Y_j}^2$ is equivalent, under $n_X/n_Y = a$, to $\sigma_{\gamma_j}^2/\sigma_{\Gamma_j}^2 = ac$ for $j=1,\dots,L$. Thus, the variance ratio $\sigma_{Y_j}^2/\sigma_{X_j}^2 = 1/c$ is constant across instruments, corresponding to the simplifying assumption in \citet[p. 1751]{10.1214/19-AOS1866}.
\end{remark}

Combining the exact variance ordering established above with the fixed-$L$ variance orders yields the asymptotic equivalence of the IVW and PS variances.

\begin{corollary}[Fixed-$L$ asymptotic variance equivalence]
    Suppose $n_X/n_Y = a$,  $L$ is fixed as $n := \min(n_X,n_Y) \to \infty$, and for some constant $c > 0$,
    $$\sigma_{X_j}^2 = c\sigma_{Y_j}^2, \qquad j = 1,\dots,L.$$
    Then the relative variance gap satisfies
    $$0 < \frac{V_{\mathtt{PS}}}{V_{\mathtt{IVW}}}-1 = \frac{V_{\mathtt{PS}}^{(2)}}{V_{\mathtt{IVW}}} = \Theta(n^{-1}).$$
    In particular,
    $$\lim\limits_{n\to\infty}\frac{V_{\mathtt{PS}}}{V_{\mathtt{IVW}}} = 1.$$
    Since $V_{\mathtt{dIVW}} = V_{\mathtt{IVW}}$ to first order for fixed $L$, the same asymptotic equivalence holds for dIVW and PS.
\end{corollary}

\begin{proof}
By the preceding proposition and Lemma \ref{lem:exact-variance-expressions},
$$\frac{V_{\mathtt{PS}}}{V_{\mathtt{IVW}}}-1=\frac{V_{\mathtt{PS}}^{(2)}}{V_{\mathtt{IVW}}} = \frac{cL}{(c\beta^2+1)\sum_{j=1}^L \gamma_j^2/\sigma_{Y_j}^2} = \Theta(n^{-1}) \to 0.$$
Therefore, $$\lim\limits_{n\to\infty}\frac{V_{\mathtt{PS}}}{V_{\mathtt{IVW}}} = 1.$$
\end{proof}

Thus, under the common variance-ratio condition, PS and IVW/dIVW have the same first-order asymptotic variance for fixed $L$, while the PS variance formula contains an additional positive term of order $n^{-2}$. The positive difference is precisely the profile-score information-bias contribution in this diagonal summary-data model. This additional component is of order $n^{-2}$ and is therefore negligible relative to the order-$n^{-1}$ leading variance. We therefore interpret this result as a comparison of lower-order terms in the asymptotic variance formulae, rather than a higher-order comparison of the estimators' actual sampling variances.

This asymptotic equivalence relies on the fixed-$L$ setting, in which average instrument strength $\kappa$ diverges as the sample sizes increase. When $L\to\infty$, however, $\kappa$ need not diverge, and the additional PS variance component may no longer be negligible relative to the leading variance. Moreover, the IVW and dIVW variance formulae differ in this regime because the uncertainty in the estimated first-stage associations can remain consequential. We therefore extend the comparison to diverging $L$, allowing $\kappa$ to vary with the number of instruments.

\subsection{Diverging number of instruments}\label{sec:diverging-L-variance}
We now establish the asymptotic orders of the IVW, dIVW, and PS variance formulae as $L\to\infty$, allowing for different regimes of average instrument strength. Under a common variance-ratio condition, we derive exact expressions and show that the PS variance formula lies strictly above the IVW variance formula and at or below the dIVW variance formula. We further establish that the absolute variance gaps vanish when $\sqrt{L}\kappa\to\infty$, whereas all three variance formulae are relatively asymptotically equivalent if and only if $\kappa\to\infty$. Throughout this subsection, the comparisons concern the variance formulae themselves, independently of the conditions required for asymptotic normality of the corresponding estimators around the true causal effect.

In particular, the comparisons involving $\mathsf{V}_{\mathtt{IVW}}$ outside the strong-instrument normality regime are algebraic comparisons of the specified variance formulae. They do not, by themselves, establish that these expressions represent the asymptotic sampling variances of the corresponding estimators under every instrument-strength sequence.

\subsubsection{Variance decomposition and asymptotic orders}
To establish this comparison, we first recall the variance expressions established in Lemmas \ref{nom1} and \ref{nom2}, Theorem \ref{asymnorm}, Lemmas \ref{lem:2} and \ref{lem:3}, and Theorem \ref{thm:clt-simple}.
$$\mathcal{V}_1 = \sum\limits_{j=1}^L\frac{\gamma_j^2}{\sigma_{Y_j}^2}\left(1+\beta^2\frac{\sigma_{X_j}^2}{\sigma_{Y_j}^2}\right) \asymp L\kappa ~\mathrm{and}~ \mathcal{V}_2 = \sum\limits_{j=1}^L\frac{\sigma_{X_j}^2}{\sigma_{Y_j}^2}\left(1+2\beta^2\frac{\sigma_{X_j}^2}{\sigma_{Y_j}^2}\right) \asymp L.$$
$$\mathsf{V}_{\mathtt{IVW}} = \frac{\mathcal{V}_1+\mathcal{V}_2}{\left(\sum_{j=1}^L(\gamma_j^2+\sigma_{X_j}^2)/\sigma_{Y_j}^2\right)^2}.$$
$$\mathsf{V}_{\mathtt{dIVW}} = \frac{\mathcal{V}_1+\mathcal{V}_2}{\left(\sum_{j=1}^L\gamma_j^2/\sigma_{Y_j}^2\right)^2}.$$
$$\tilde{V}_1 = \sum_{j=1}^L \frac{\gamma_j^2 \sigma_{Y_j}^2 + \Gamma_j^2 \sigma_{X_j}^2 + \sigma_{X_j}^2 \sigma_{Y_j}^2}{(\sigma_{X_j}^2 \beta^2 + \sigma_{Y_j}^2)^2}.$$
$$\tilde{V}_2 = \tilde{V}_1 - \sum\limits_{j=1}^L \frac{\sigma_{X_j}^2 \sigma_{Y_j}^2}{(\sigma_{X_j}^2 \beta^2 + \sigma_{Y_j}^2)^2} = \sum\limits_{j=1}^L \frac{\gamma_j^2 \sigma_{Y_j}^2 + \Gamma_j^2 \sigma_{X_j}^2}{(\sigma_{X_j}^2 \beta^2 + \sigma_{Y_j}^2)^2} = \sum\limits_{j=1}^L\frac{\gamma_j^2}{\sigma_{X_j}^2 \beta^2 + \sigma_{Y_j}^2}.$$
$$V_{\mathtt{PS}} = \frac{\tilde{V}_1}{\tilde{V}_2^2}.$$
Since $\sigma_{X_j}^2 > 0$, $\mathsf{V}_{\mathtt{IVW}} < \mathsf{V}_{\mathtt{dIVW}}$.
The decomposition in Lemma \ref{lem:PS-var-decomp} continues to hold for diverging $L$, since it is an algebraic identity for every $L$.
\begin{align*}
V_{\mathtt{PS}}
&= V_{\mathtt{PS}}^{(1)}+V_{\mathtt{PS}}^{(2)},\\
V_{\mathtt{PS}}^{(1)}
&= \left(\sum_{j=1}^L
\frac{\gamma_j^2}
{\sigma_{X_j}^2\beta^2+\sigma_{Y_j}^2}
\right)^{-1},\\
V_{\mathtt{PS}}^{(2)}
&= \frac{1}{\tilde V_2^2}
\sum_{j=1}^L
\frac{\sigma_{X_j}^2\sigma_{Y_j}^2}
{(\sigma_{X_j}^2\beta^2+\sigma_{Y_j}^2)^2}.
\end{align*}

Before imposing additional conditions to compare the variances exactly, we first establish the asymptotic orders of the PS variance components and the IVW and dIVW variances in the diverging-$L$ setting, allowing the average instrument strength $\kappa$ to vary with $L$.
\begin{proposition}[Diverging-$L$ variance orders]\label{prop:diverging-L-variance-orders}
    Suppose $n_X/n_Y \to a \in (0,\infty)$ and $L \to \infty$ as $n := \min(n_X,n_Y) \to \infty$. Assume that $\kappa > 0$ and $\sigma_{X_j}^2 \asymp \sigma_{Y_j}^2$ uniformly over $j = 1,\dots,L$.
    Then, $$V_{\mathtt{PS}}^{(1)} = \Theta\left(\frac{1}{L\kappa}\right), \qquad V_{\mathtt{PS}}^{(2)} = \Theta\left(\frac{1}{L\kappa^2}\right)$$
    and hence
    $$V_{\mathtt{PS}} = \Theta\left(\frac{\kappa+1}{L\kappa^2}\right).$$
    Moreover,
    $$\mathsf{V}_{\mathtt{IVW}} = \Theta\left(\frac{1}{L(\kappa+1)}\right), \qquad \mathsf{V}_{\mathtt{dIVW}} = \Theta\left(\frac{\kappa+1}{L\kappa^2}\right).$$
    Consequently,
    $$\frac{V_{\mathtt{PS}}^{(2)}}{V_{\mathtt{PS}}^{(1)}} = \Theta(\kappa^{-1}), \qquad \frac{V_{\mathtt{PS}}^{(2)}}{\mathsf{V}_{\mathtt{IVW}}} = \Theta\left(\frac{\kappa+1}{\kappa^2}\right), \qquad \frac{V_{\mathtt{PS}}^{(2)}}{\mathsf{V}_{\mathtt{dIVW}}} =\Theta\left(\frac{1}{\kappa+1}\right).$$
\end{proposition}

The proof is provided in Appendix \ref{app:proofs}. In particular, the variance orders imply
$$\frac{\mathsf{V}_{\mathtt{IVW}}}{V_{\mathtt{PS}}} = \Theta\left(\frac{\kappa^2}{(\kappa+1)^2}\right).$$
Thus, IVW has a strictly smaller asymptotic variance order than PS when $\kappa\to0$, whereas the two variances have the same asymptotic order when $\kappa$ remains bounded away from zero.

The PS and dIVW variances have the same asymptotic order across all instrument-strength regimes. However, the smaller variance order of IVW when $\kappa \to 0$ should not be interpreted as an efficiency advantage, since IVW may exhibit substantial weak-instrument bias. The variance comparisons alone therefore do not establish an ordering of mean squared errors or inferential performance. Moreover, the additional PS variance component is asymptotically negligible relative to the leading PS component and both the IVW and dIVW variances when $\kappa \to \infty$. When $\kappa$ remains bounded away from zero and infinity, however, the additional component is of the same asymptotic order as these variances. Although the requirement $\kappa \to \infty$ is implied by the IVW condition ($\kappa/L \to \infty$), it is not required for dIVW ($\sqrt{L}\kappa \to \infty$) or PS (for which asymptotic normality can be established under either $L \to \infty$ with $\|\boldsymbol{\gamma}\|_3/\|\boldsymbol{\gamma}\|_2 \to 0$, or $\kappa \to \infty$, together with the underlying regularity conditions).

\subsubsection{Exact variance ordering and asymptotic equivalence}
Having established the general variance orders, we now impose a common variance-ratio condition to obtain exact expressions for the three variance formulae. These expressions allow us to establish their ordering and distinguish the conditions for absolute variance convergence from those for relative asymptotic equivalence.
\begin{lemma}\label{lem:exact-variance-expressions-diverging}
    Suppose that, for some constant $c > 0$,
    $$\sigma_{X_j}^2 = c\sigma_{Y_j}^2, \qquad j = 1,\dots,L.$$
    Then, for every $L \geq 1$,
    $$V_{\mathtt{PS}} = V_{\mathtt{PS}}^{(1)}+V_{\mathtt{PS}}^{(2)} = \frac{1+c\beta^2}{cL\kappa}+\frac{1}{cL\kappa^2},$$
    $$\mathsf{V}_{\mathtt{IVW}} = \frac{1+c\beta^2}{cL(\kappa+1)}+\frac{\beta^2}{L(\kappa+1)^2}, \qquad \text{ and } \qquad \mathsf{V}_{\mathtt{dIVW}} = \frac{1+c\beta^2}{cL\kappa}+\frac{1+2c\beta^2}{cL\kappa^2}.$$
\end{lemma}

The proof is provided in Appendix \ref{app:proofs}.

This yields an exact ordering between the IVW, PS and dIVW variances.
\begin{theorem}[Exact variance ordering]\label{thm:diverging-L-variance-ordering}
Suppose that, for some constant $c > 0$,
$$\sigma_{X_j}^2 = c\sigma_{Y_j}^2, \qquad j = 1,\dots,L.$$
Then, for every $L \geq 1$,
$$V_{\mathtt{PS}}^{(1)}-\mathsf{V}_{\mathtt{IVW}} = \frac{\kappa+1+c\beta^2}{cL\kappa(\kappa+1)^2} > 0$$
and
$$\mathsf{V}_{\mathtt{dIVW}}-V_{\mathtt{PS}} = \frac{2\beta^2}{L\kappa^2} \geq 0.$$
Therefore, $\mathsf{V}_{\mathtt{IVW}} < V_{\mathtt{PS}} \leq \mathsf{V}_{\mathtt{dIVW}}$, where equality in the second inequality holds if and only if $\beta=0$.
\end{theorem}

\begin{proof}
By Lemma \ref{lem:exact-variance-expressions-diverging},
\begin{align*}
V_{\mathtt{PS}}^{(1)}-\mathsf{V}_{\mathtt{IVW}} &= \frac{1+c\beta^2}{cL\kappa}-\frac{1+c\beta^2}{cL(\kappa+1)}-\frac{\beta^2}{L(\kappa+1)^2} \\
&= \frac{1+c\beta^2}{cL}\frac{1}{\kappa(\kappa+1)}-\frac{\beta^2}{L(\kappa+1)^2} \\
&= \frac{1}{L(\kappa+1)}\frac{(1+c\beta^2)(\kappa+1)-c\beta^2\kappa}{c\kappa(\kappa+1)} \\
&= \frac{1}{L(\kappa+1)}\frac{\kappa+1+c\beta^2}{c\kappa(\kappa+1)} \\
&= \frac{\kappa+1+c\beta^2}{cL\kappa(\kappa+1)^2} > 0, \\
\mathsf{V}_{\mathtt{dIVW}}-V_{\mathtt{PS}} &= \frac{2c\beta^2}{cL\kappa^2} = \frac{2\beta^2}{L\kappa^2} \geq 0,
\end{align*}
since $\kappa > 0$, $c > 0$, $\beta^2 \geq 0$, $L \geq 1$. Since $V_{\mathtt{PS}}^{(2)} > 0$,
$$\mathsf{V}_{\mathtt{IVW}} < V_{\mathtt{PS}}^{(1)} < V_{\mathtt{PS}}^{(1)}+V_{\mathtt{PS}}^{(2)} = V_{\mathtt{PS}} \leq \mathsf{V}_{\mathtt{dIVW}}.$$
\end{proof}

\begin{remark}
    The constant variance ratio $\sigma_{Y_j}^2/\sigma_{X_j}^2 = 1/c$ across instruments, corresponds to the simplifying assumption in \citet[p. 1751]{10.1214/19-AOS1866}.
\end{remark}

Under the common variance ratio condition, if $L\kappa^2\to\infty$, the absolute differences among the PS, IVW and dIVW variances converge to zero. However, this condition does not imply their relative asymptotic equivalence. In particular, the relative variance gaps are $O(\kappa^{-1})$ as $\kappa\to\infty$, under which all three variances are asymptotically equivalent.

\begin{proposition}[Diverging-$L$ variance equivalence]\label{prop:diverging-L-variance-equivalence}
    Suppose $L \to \infty$ as $n := \min(n_X,n_Y)$
    $ \to \infty$, and, for some constant $c > 0$,
    $$\sigma_{X_j}^2 = c\sigma_{Y_j}^2, \qquad j = 1,\dots,L.$$
    Then, the absolute variance gaps satisfy
    \begin{align*}
    V_{\mathtt{PS}}-\mathsf{V}_{\mathtt{IVW}} &= \frac{\kappa+1+c\beta^2}{cL\kappa(\kappa+1)^2}+\frac{1}{cL\kappa^2} = \Theta\left(\frac{1}{L\kappa^2}\right) \\
    \mathsf{V}_{\mathtt{dIVW}}-V_{\mathtt{PS}} &= \frac{2\beta^2}{L\kappa^2}.
    \end{align*}
    The first gap vanishes if and only if $\sqrt{L}\kappa \to \infty$. The same equivalence holds for the second gap when $\beta \neq 0$, whereas the second gap is identically zero when $\beta = 0$.
    
    Moreover, the relative variance gaps satisfy
    \begin{align*}
    \frac{V_{\mathtt{PS}}}{\mathsf{V}_{\mathtt{IVW}}}-1 &= \frac{2\kappa^2+(3+c\beta^2)\kappa+1}{(1+c\beta^2)\kappa^3+(1+2c\beta^2)\kappa^2} \\
    \frac{\mathsf{V}_{\mathtt{dIVW}}}{V_{\mathtt{PS}}}-1 &= \frac{2c\beta^2}{(1+c\beta^2)\kappa+1}
    \end{align*}
    Therefore,
    $$\lim\limits_{n\to\infty}\frac{V_{\mathtt{PS}}}{\mathsf{V}_{\mathtt{IVW}}} = \lim\limits_{n\to\infty}\frac{\mathsf{V}_{\mathtt{dIVW}}}{V_{\mathtt{PS}}} = 1 \quad \Longleftrightarrow \quad \kappa \to \infty.$$
\end{proposition}

\begin{proof}
By the preceding theorem and Lemma \ref{lem:exact-variance-expressions-diverging},
\begin{align*}
V_{\mathtt{PS}}-\mathsf{V}_{\mathtt{IVW}} &= (V_{\mathtt{PS}}^{(1)}-\mathsf{V}_{\mathtt{IVW}})+V_{\mathtt{PS}}^{(2)} \\
&= \frac{\kappa+1+c\beta^2}{cL\kappa(\kappa+1)^2}+\frac{1}{cL\kappa^2}, \\
\mathsf{V}_{\mathtt{dIVW}}-V_{\mathtt{PS}} &= \frac{2\beta^2}{L\kappa^2}.
\end{align*}
Since $\kappa+1 > \kappa$ and $\kappa+1 > 1$,
\begin{align*}
\frac{\kappa+1+c\beta^2}{cL\kappa(\kappa+1)^2} &= \frac{1}{cL\kappa(\kappa+1)}+\frac{\beta^2}{L\kappa(\kappa+1)^2} \\
&< \frac{1}{cL\kappa^2}+\frac{\beta^2}{L\kappa^2} = \frac{1+c\beta^2}{L\kappa^2}.
\end{align*}
Consequently,
$$\frac{1}{cL\kappa^2} \leq V_{\mathtt{PS}}-\mathsf{V}_{\mathtt{IVW}} < \frac{2+c\beta^2}{cL\kappa^2},$$
which establishes
$$V_{\mathtt{PS}}-\mathsf{V}_{\mathtt{IVW}} = \Theta\left(\frac{1}{L\kappa^2}\right).$$
Therefore, the first absolute variance gap converges to zero if and only if $L\kappa^2\to\infty$, equivalently $\sqrt{L}\kappa\to\infty$.
The same condition is necessary and sufficient for the second gap when $\beta\neq0$, while the second gap is identically zero when $\beta=0$.

Next, dividing the first absolute variance gap by $\mathsf{V}_{\mathtt{IVW}}$ yields
\begin{align*}
\frac{V_{\mathtt{PS}}}{\mathsf{V}_{\mathtt{IVW}}}-1
&= \frac{(\kappa+1+c\beta^2)\kappa+(\kappa+1)^2}
{\kappa^2[(1+c\beta^2)(\kappa+1)+c\beta^2]} \\
&= \frac{2\kappa^2+(3+c\beta^2)\kappa+1}
{(1+c\beta^2)\kappa^3+(1+2c\beta^2)\kappa^2}.
\end{align*}
For fixed $c>0$ and $\beta$, dividing the numerator and denominator by $\kappa^3$ gives
\begin{align*}
\frac{V_{\mathtt{PS}}}{\mathsf{V}_{\mathtt{IVW}}}-1
&= \frac{2\kappa^{-1}+(3+c\beta^2)\kappa^{-2}+\kappa^{-3}}{1+c\beta^2+(1+2c\beta^2)\kappa^{-1}} \\
&= \frac{2}{(1+c\beta^2)\kappa}+O(\kappa^{-2}),
\end{align*}
as $\kappa\to\infty$.
The second relative gap
$$\frac{\mathsf{V}_{\mathtt{dIVW}}}{V_{\mathtt{PS}}}-1 = \frac{2c\beta^2}{(1+c\beta^2)\kappa+1} = \frac{2c\beta^2}{(1+c\beta^2)\kappa}+O(\kappa^{-2})$$
also converges to zero as $\kappa\to\infty$.
Consequently,
$$\lim_{n\to\infty} \frac{V_{\mathtt{PS}}}{\mathsf{V}_{\mathtt{IVW}}}
= \lim_{n\to\infty}\frac{\mathsf{V}_{\mathtt{dIVW}}}{V_{\mathtt{PS}}} =1.$$
Conversely, the first relative variance gap is strictly positive for finite $\kappa>0$ and diverges as $\kappa\to0$. Hence, relative asymptotic equivalence of all three variance formulae holds if and only if $\kappa\to\infty$.
\end{proof}

\section{Discussion}
The comparisons reveal a sharp distinction between the fixed- and many-instrument regimes. With a fixed number of instruments, the standard two-sample IV limits imply that IVW and dIVW are consistent and asymptotically normal and have the same first-order asymptotic variance; the dIVW bias correction is of smaller order than the first-order stochastic variation. Under a common variance-ratio condition, the PS variance formula equals the IVW variance plus a positive information-bias component. That component is $\Theta(n^{-2})$, compared with the $\Theta(n^{-1})$ leading variance, so the relative gap is $\Theta(n^{-1})$. Hence PS and IVW/dIVW have the same first-order asymptotic variance, although the PS variance formula contains an additional lower-order term.

When the number of instruments diverges, the estimators require substantially different strength conditions. IVW is consistent when $\kappa\to\infty$, while its centred asymptotic normal approximation requires the stronger condition $\kappa/L\to\infty$. By contrast, dIVW is consistent and asymptotically normal under $\sqrt{L}\kappa\to\infty$, together with the no-dominating-instrument condition. Under the variance assumptions used for PS, its collective instrument strength condition $L/(n^2\|\boldsymbol{\gamma}\|_2^4)\to0$ is equivalent to $\sqrt{L}\kappa\to\infty$, although PS and dIVW impose different additional restrictions on the distribution of instrument strength.

The diverging-$L$ variance comparisons further show that the PS and dIVW variance formulae have the same asymptotic order across instrument-strength regimes, whereas the IVW variance formula has a smaller order when $\kappa\to0$. Under a common variance-ratio condition, the exact ordering $\mathsf{V}_{\mathtt{IVW}}<V_{\mathtt{PS}}\leq\mathsf{V}_{\mathtt{dIVW}}$ holds, with equality in the second inequality if and only if $\beta=0$. The absolute variance gaps vanish when $\sqrt{L}\kappa\to\infty$, whereas relative asymptotic equivalence of all three variance formulae holds if and only if $\kappa\to\infty$. Thus, the collective-strength condition used to establish dIVW consistency and asymptotic normality also ensures vanishing absolute variance gaps, but does not necessarily imply relative variance equivalence. The smaller IVW variance formula under weak instruments should not be interpreted as an efficiency advantage, since IVW may exhibit substantial weak-instrument bias.

These results clarify why weak-instrument corrections matter in two-sample IV with many instruments and why conclusions from the fixed-instrument regime do not directly transfer to high-dimensional settings. They also show that comparisons between dIVW and PS depend not only on average instrument strength but on how that strength is distributed across instruments and on instrument-set growth relative to sample size. Although the dIVW asymptotic variance formula exceeds the PS variance formula under the common variance-ratio condition when $\beta\neq0$, dIVW and PS satisfy different sufficient conditions for asymptotic normality. Their relative merits cannot be determined from variance comparisons alone. These conclusions apply to the independent summary-association model studied here; extending the comparison to correlated instruments requires a matrix-valued covariance formulation and is not implied by the present proofs.

\begingroup
\setlength{\bibsep}{2.75pt}
\bibliographystyle{ecta}
\bibliography{refs}
\endgroup

\section{Additional Proofs}\label{app:proofs}
\begin{proof}[Proof of Proposition \ref{consis-Lfix}]
From the previous section, we know that $\hat{\gamma}_j \overset{p}{\to} \gamma_j$, $\hat{\Gamma}_j \overset{p}{\to} \Gamma_j$, and $n_Y\hat{\sigma}_{Y_j}^2 \overset{p}{\rightarrow} \sigma_{\Gamma_j}^2$. We can write
$$\hat{\beta}_{\mathtt{IVW}} = \frac{\sum_{j=1}^L\hat{\Gamma}_j\hat{\gamma}_j/\hat{\sigma}_{Y_j}^2}{\sum_{j=1}^L\hat{\gamma}_j^2/\hat{\sigma}_{Y_j}^2} = \frac{\sum_{j=1}^L\hat{\Gamma}_j\hat{\gamma}_j/n_Y\hat{\sigma}_{Y_j}^2}{\sum_{j=1}^L\hat{\gamma}_j^2/n_Y\hat{\sigma}_{Y_j}^2}.$$
Since $L$ is fixed, these componentwise convergences and Slutsky's theorem imply

$\sum_{j=1}^L\hat{\gamma}_j^2/n_Y\hat{\sigma}_{Y_j}^2 \overset{p}{\rightarrow} \sum_{j=1}^L\gamma_j^2/\sigma_{\Gamma_j}^2$ and $\sum_{j=1}^L\hat{\Gamma}_j\hat{\gamma}_j/n_Y\hat{\sigma}_{Y_j}^2 \overset{p}{\rightarrow} \sum_{j=1}^L\Gamma_j\gamma_j/\sigma_{\Gamma_j}^2$.

Since $\gamma_j$'s are not all zero by Assumption \ref{assume:structure}, the limiting denominator is strictly positive. Dividing by the denominator and using Slutsky's theorem gives
\begin{align*}
\hat{\beta}_{\mathtt{IVW}} \overset{p}{\to} \frac{\sum_{j=1}^L\Gamma_j\gamma_j/\sigma_{\Gamma_j}^2}{\sum_{j=1}^L\gamma_j^2/\sigma_{\Gamma_j}^2} &= \frac{\sum_{j=1}^L\beta\gamma_j^2/\sigma_{\Gamma_j}^2}{\sum_{j=1}^L\gamma_j^2/\sigma_{\Gamma_j}^2} = \beta,
\end{align*}
where the first equality follows from Assumption \ref{assume:structure}.
\end{proof}

\begin{proof}[Proof of Proposition \ref{asymnorm-Lfix}]
We write
\begin{align*}
\hat{\beta}_{\mathtt{IVW}} &= \frac{\sum_{j=1}^L[\hat{\gamma}_j\beta-(\hat{\gamma}_j-\gamma_j)\beta+(\hat{\Gamma}_j-\Gamma_j)]\hat{\gamma}_j/n_Y\hat{\sigma}_{Y_j}^2}{\sum_{j=1}^L\hat{\gamma}_j^2/n_Y\hat{\sigma}_{Y_j}^2} \\
&= \beta+\frac{\sum_{j=1}^L(\hat{\gamma}_j/n_Y\hat{\sigma}_{Y_j}^2)[-(\hat{\gamma}_j-\gamma_j)\beta+
(\hat{\Gamma}_j-\Gamma_j)]}{\sum_{j=1}^L\hat{\gamma}_j^2/n_Y\hat{\sigma}_{Y_j}^2}
\end{align*}
As $n \to \infty$, $n_X/n_Y \to a$ and
\begin{align*}
\sqrt{n_Y}(\hat{\Gamma}_j-\Gamma_j) &\overset{d}{\rightarrow} \mathcal{N}(0,\sigma_{\Gamma_j}^2), \\
\sqrt{n_X}(\hat{\gamma}_j-\gamma_j) &\overset{d}{\rightarrow} \mathcal{N}(0,\sigma_{\gamma_j}^2).
\end{align*}
For the first component of the numerator,
\begin{align*}
-(\hat{\gamma}_j-\gamma_j)\beta &= -\frac{\sqrt{n_Y}\sqrt{n_X}}{\sqrt{n_Y}\sqrt{n_X}}(\hat{\gamma}_j-\gamma_j)\beta \\
&= -\frac{1}{\sqrt{n_Y}}\frac{\sqrt{n_Y}}{\sqrt{n_X}}\sqrt{n_X}(\hat{\gamma}_j-\gamma_j)\beta.
\end{align*}
Since $\sqrt{n_X}(\hat{\gamma}_j-\gamma_j) \overset{d}{\rightarrow} \mathcal{N}(0,\sigma_{\gamma_j}^2)$, $\sqrt{n_X}(\hat{\gamma}_j-\gamma_j)\beta \overset{d}{\to} \mathcal{N}(0,\beta^2\sigma_{\gamma_j}^2)$.

We also have $\sqrt{n_Y}/\sqrt{n_X} \to 1/\sqrt{a}$. By Slutsky's theorem,
$$-\frac{\sqrt{n_Y}}{\sqrt{n_X}}\sqrt{n_X}(\hat{\gamma}_j-\gamma_j)\beta \overset{d}{\rightarrow} \mathcal{N}\left(0,\frac{\beta^2\sigma_{\gamma_j}^2}{a}\right).$$
For the second component,
\begin{align*}
\hat{\Gamma}_j-\Gamma_j &= \frac{1}{\sqrt{n_Y}}\sqrt{n_Y}(\hat{\Gamma}_j-\Gamma_j)
\end{align*}
and we know $\sqrt{n_Y}(\hat{\Gamma}_j-\Gamma_j) \overset{d}{\rightarrow} \mathcal{N}(0,\sigma_{\Gamma_j}^2)$.

By independence of the exposure and outcome estimation errors,
$$-\sqrt{n_Y}(\hat{\gamma}_j-\gamma_j)\beta+\sqrt{n_Y}(\hat{\Gamma}_j-\Gamma_j) \overset{d}{\to} \mathcal{N}\left(0,\frac{\beta^2\sigma_{\gamma_j}^2}{a}+\sigma_{\Gamma_j}^2\right).$$
Since $\hat{\gamma}_j \overset{p}{\to} \gamma_j$ and $n_Y\hat{\sigma}_{Y_j}^2 \overset{p}{\rightarrow} \sigma_{\Gamma_j}^2$, $\hat{\gamma}_j/n_Y\hat{\sigma}_{Y_j}^2 \overset{p}{\rightarrow} \gamma_j/\sigma_{\Gamma_j}^2$. By Slutsky's theorem,
\begin{align*}
\frac{\hat{\gamma}_j}{n_Y\hat{\sigma}_{Y_j}^2}\left[-\sqrt{n_Y}(\hat{\gamma}_j-\gamma_j)\beta+\sqrt{n_Y}(\hat{\Gamma}_j-\Gamma_j)\right] &\overset{d}{\rightarrow} \mathcal{N}\left(0,\frac{\gamma_j^2\beta^2\sigma_{\gamma_j}^2}{a(\sigma_{\Gamma_j}^2)^2}+\frac{\gamma_j^2\sigma_{\Gamma_j}^2}{(\sigma_{\Gamma_j}^2)^2}\right) \\
&= \mathcal{N}\left(0,\frac{1}{a}\frac{\gamma_j^2}{\sigma_{\Gamma_j}^2}\beta^2\frac{\sigma_{\gamma_j}^2}{\sigma_{\Gamma_j}^2}+\frac{\gamma_j^2}{\sigma_{\Gamma_j}^2}\right).
\end{align*}
Since $L$ is fixed, independence across instruments gives
$$\sum\limits_{j=1}^L\frac{\hat{\gamma}_j}{n_Y\hat{\sigma}_{Y_j}^2}\left[-\sqrt{n_Y}(\hat{\gamma}_j-\gamma_j)\beta+\sqrt{n_Y}(\hat{\Gamma}_j-\Gamma_j)\right] \overset{d}{\rightarrow} \mathcal{N}\left(0,\frac{1}{a}\sum\limits_{j=1}^L\frac{\gamma_j^2}{\sigma_{\Gamma_j}^2}\beta^2\frac{\sigma_{\gamma_j}^2}{\sigma_{\Gamma_j}^2}+\sum\limits_{j=1}^L\frac{\gamma_j^2}{\sigma_{\Gamma_j}^2}\right).$$
Equivalently, $$\sqrt{n_Y}\sum\limits_{j=1}^L\frac{\hat{\gamma}_j}{n_Y\hat{\sigma}_{Y_j}^2}\left[-(\hat{\gamma}_j-\gamma_j)\beta+(\hat{\Gamma}_j-\Gamma_j)\right] \overset{d}{\to} \mathcal{N}\left(0,n_Y(V_1+V_2)\right).$$
From Proof 1 of Proposition \ref{consis-Lfix}, $\sum_{j=1}^L\hat{\gamma}_j^2/n_Y\hat{\sigma}_{Y_j}^2 \overset{p}{\rightarrow} \sum_{j=1}^L\gamma_j^2/\sigma_{\Gamma_j}^2$, where the limiting denominator is strictly positive by Assumption \ref{assume:structure}. Hence, dividing by the denominator and using Slutsky's theorem gives
$$\sqrt{n_Y}(\hat{\beta}_{\mathtt{IVW}}-\beta) \overset{d}{\to} \mathcal{N}(0, n_YV_{\mathtt{IVW}}).$$
Therefore,
$$\frac{1}{\sqrt{V_{\mathtt{IVW}}}}(\hat{\beta}_{\mathtt{IVW}}-\beta) \overset{d}{\to} \mathcal{N}(0,1),$$
which establishes the result.
\end{proof}

\begin{proof}[Proof of Proposition \ref{consis-Lfix-dIVW}] We write
$$\hat{\beta}_{\mathtt{dIVW}} = \frac{\sum\limits_{j=1}^L\hat{\Gamma}_j\hat{\gamma}_j/\hat{\sigma}_{Y_j}^2}{\sum\limits_{j=1}^L(\hat{\gamma}_j^2-\hat{\sigma}_{X_j}^2)/\hat{\sigma}_{Y_j}^2} = \frac{\sum\limits_{j=1}^L\hat{\Gamma}_j\hat{\gamma}_j/n_Y\hat{\sigma}_{Y_j}^2}{\sum\limits_{j=1}^L(\hat{\gamma}_j^2-\hat{\sigma}_{X_j}^2)/n_Y\hat{\sigma}_{Y_j}^2}.$$
From the proof of Proposition \ref{consis-Lfix},
\begin{align*}
&\sum\limits_{j=1}^L\frac{\hat{\Gamma}_j\hat{\gamma}_j}{n_Y\hat{\sigma}_{Y_j}^2} \overset{p}{\rightarrow} \sum\limits_{j=1}^L\frac{\Gamma_j\gamma_j}{\sigma_{\Gamma_j}^2}, \\
&\frac{\hat{\gamma}_j^2}{n_Y\hat{\sigma}_{Y_j}^2} \overset{p}{\rightarrow} \frac{\gamma_j^2}{\sigma_{\Gamma_j}^2}.
\end{align*}
We have
$$\frac{\hat{\sigma}_{X_j}^2}{n_Y\hat{\sigma}_{Y_j}^2} = \frac{1}{n_X}\frac{n_X\hat{\sigma}_{X_j}^2}{n_Y\hat{\sigma}_{Y_j}^2} \overset{p}{\to} 0 \cdot \frac{\sigma_{\gamma_j}^2}{\sigma_{\Gamma_j}^2} = 0.$$
Then by Slutsky's theorem, $(\hat{\gamma}_j^2-\hat{\sigma}_{X_j}^2)/n_Y\hat{\sigma}_{Y_j}^2 \overset{p}{\to}  \gamma_j^2/\sigma_{\Gamma_j}^2$ so $\sum_{j=1}^L(\hat{\gamma}_j^2-\hat{\sigma}_{X_j}^2)/n_Y\hat{\sigma}_{Y_j}^2 \overset{p}{\rightarrow} \sum_{j=1}^L\gamma_j^2/\sigma_{\Gamma_j}^2$. Finally,
\begin{align*}
\hat{\beta}_{\mathtt{dIVW}} \overset{p}{\to} \frac{\sum\limits_{j=1}^L\Gamma_j\gamma_j/\sigma_{\Gamma_j}^2}{\sum\limits_{j=1}^L\gamma_j^2/\sigma_{\Gamma_j}^2} &= \frac{\sum\limits_{j=1}^L\beta\gamma_j^2/\sigma_{\Gamma_j}^2}{\sum\limits_{j=1}^L\gamma_j^2/\sigma_{\Gamma_j}^2} = \beta.
\end{align*}
\end{proof}

\begin{proof}[Proof of Proposition \ref{asymnorm-Lfix-dIVW}]
We write
\begin{align*}
\hat{\beta}_{\mathtt{dIVW}} &= \frac{\sum\limits_{j=1}^L[\hat{\gamma}_j\beta-(\hat{\gamma}_j-\gamma_j)\beta+(\hat{\Gamma}_j-\Gamma_j)]\hat{\gamma}_j/n_Y\hat{\sigma}_{Y_j}^2}{\sum\limits_{j=1}^L(\hat{\gamma}_j^2-\hat{\sigma}_{X_j}^2)/n_Y\hat{\sigma}_{Y_j}^2} \\
&= \frac{\sum\limits_{j=1}^L\left\{(\hat{\gamma}_j^2-\hat{\sigma}_{X_j}^2+\hat{\sigma}_{X_j}^2)\beta/n_Y\hat{\sigma}_{Y_j}^2+(\hat{\gamma}_j/n_Y\hat{\sigma}_{Y_j}^2)[-(\hat{\gamma}_j-\gamma_j)\beta+
(\hat{\Gamma}_j-\Gamma_j)]\right\}}{\sum\limits_{j=1}^L(\hat{\gamma}_j^2-\hat{\sigma}_{X_j}^2)/n_Y\hat{\sigma}_{Y_j}^2} \\
&= \beta+\frac{\sum\limits_{j=1}^L\left\{\beta\hat{\sigma}_{X_j}^2/n_Y\hat{\sigma}_{Y_j}^2+(\hat{\gamma}_j/n_Y\hat{\sigma}_{Y_j}^2)[-(\hat{\gamma}_j-\gamma_j)\beta+
(\hat{\Gamma}_j-\Gamma_j)]\right\}}{\sum\limits_{j=1}^L(\hat{\gamma}_j^2-\hat{\sigma}_{X_j}^2)/n_Y\hat{\sigma}_{Y_j}^2}
\end{align*}
By the argument used in the proof of Proposition \ref{consis-Lfix-dIVW},
$$\sum\limits_{j=1}^L\frac{\hat{\gamma}_j^2-\hat{\sigma}_{X_j}^2}{n_Y\hat{\sigma}_{Y_j}^2} \overset{p}{\rightarrow} \sum\limits_{j=1}^L\frac{\gamma_j^2}{\sigma_{\Gamma_j}^2}.$$
The limiting denominator is strictly positive because the $\gamma_j$'s are not all zero.

Next consider the additional term arising from the dIVW bias correction. For each $j$,
$$\frac{\hat{\sigma}_{X_j}^2}{n_Y\hat{\sigma}_{Y_j}^2} = \frac{1}{n_X}\frac{n_X\hat{\sigma}_{X_j}^2}{n_Y\hat{\sigma}_{Y_j}^2} = O_p(n^{-1}).$$
Since $L$ is fixed,
$$\sum\limits_{j=1}^L\frac{\hat{\sigma}_{X_j}^2}{n_Y\hat{\sigma}_{Y_j}^2} = O_p(n^{-1}) = o_p(n^{-1/2}).$$
By the argument used in the proof of Proposition \ref{asymnorm-Lfix}, the remaining numerator term
$$\sum\limits_{j=1}^L\frac{\hat{\gamma}_j}{n_Y\hat{\sigma}_{Y_j}^2}[-(\hat{\gamma}_j-\gamma_j)\beta+
(\hat{\Gamma}_j-\Gamma_j)]$$
is of order $O_p(n^{-1/2})$ and determines the first-order limiting distribution of the IVW estimator. Since the additional dIVW term is $o_p(n^{-1/2})$, it does not affect the first-order limiting distribution.

Therefore, dIVW has the same first-order asymptotic distribution as IVW. By Proposition \ref{asymnorm-Lfix} and Slutsky's theorem,
$$\frac{1}{\sqrt{V_{\mathtt{dIVW}}}}(\hat{\beta}_{\mathtt{dIVW}}-\beta) \overset{d}{\to} \mathcal{N}(0,1),$$
where $V_{\mathtt{dIVW}} = V_{\mathtt{IVW}}$.
\end{proof}

\begin{proof}[Derivation of the profile log-likelihood] For each $j$,
$$\frac{\partial l}{\partial\gamma_j} = -\frac{1}{2}\left[\frac{2(\hat{\gamma}_j-\gamma_j)(-1)}{\sigma_{X_j}^2}+\frac{2(\hat{\Gamma}_j-\gamma_j\beta)(-\beta)}{\sigma_{Y_j}^2}\right] = \frac{(\hat{\gamma}_j-\gamma_j)}{\sigma_{X_j}^2}+\frac{(\hat{\Gamma}_j-\gamma_j\beta)\beta}{\sigma_{Y_j}^2}.$$
Setting to zero gives
\begin{align*}
\sigma_{Y_j}^2(\hat{\gamma}_j-\gamma_j)+\sigma_{X_j}^2(\hat{\Gamma}_j-\gamma_j\beta)\beta &= 0 \\
(\hat{\gamma}_j\sigma_{Y_j}^2+\hat{\Gamma}_j\sigma_{X_j}^2\beta)-(\sigma_{Y_j}^2+\sigma_{X_j}^2\beta^2)\gamma_j &= 0 \\
\gamma_j &= \frac{\hat{\gamma}_j\sigma_{Y_j}^2+\hat{\Gamma}_j\sigma_{X_j}^2\beta}{\sigma_{X_j}^2\beta^2+\sigma_{Y_j}^2}
\end{align*}
Substituting $\gamma_j$ into $l(\beta,\gamma_1,\dots,\gamma_L)$,
the first term simplifies to
$$\frac{1}{\sigma_{X_j}^2}(\hat{\gamma}_j-\gamma_j)^2 = \sigma_{X_j}^2\beta^2\frac{(\hat{\Gamma}_j-\beta\hat{\gamma}_j)^2}{(\sigma_{X_j}^2\beta^2+\sigma_{Y_j}^2)^2}$$
and the second term simplifies to
$$\frac{1}{\sigma_{Y_j}^2}(\hat{\Gamma}_j-\gamma_j\beta)^2 = \sigma_{Y_j}^2\frac{(\hat{\Gamma}_j-\beta\hat{\gamma}_j)^2}{(\sigma_{X_j}^2\beta^2+\sigma_{Y_j}^2)^2}$$
Adding the two terms yields the required $l(\beta)$.
\end{proof}

\begin{proof}[Proof of Proposition \ref{prop:fixed-L-variance-orders}]
Recall that the true variances are defined by
$$\sigma_{Y_j}^2:=\frac{\sigma_{\Gamma_j}^2}{n_Y},
\qquad
\sigma_{X_j}^2:=\frac{\sigma_{\gamma_j}^2}{n_X},
\qquad j=1,\dots,L.$$
Since  $n_X/n_Y \to a \in (0,\infty)$ and $n := \min(n_X,n_Y)$, we have $n_X \asymp n_Y \asymp n$. Hence, $\sigma_{X_j}^2, \sigma_{Y_j}^2 \asymp 1/n$. Since $L$ and the structural parameters $\beta$, $\gamma_j$ do not depend on $n$, and $\sum_{j=1}^L\gamma_j^2>0$ by Assumption \ref{assume:structure}, it follows that $\sigma_{X_j}^2 \beta^2 + \sigma_{Y_j}^2 \asymp 1/n$ and so
$$\tilde{V}_2 = \sum\limits_{j=1}^L\frac{\gamma_j^2}{\sigma_{X_j}^2 \beta^2 + \sigma_{Y_j}^2} \asymp n.$$
Moreover, for each $j = 1,\dots,L$,
$$\frac{\sigma_{X_j}^2 \sigma_{Y_j}^2}{(\sigma_{X_j}^2 \beta^2 + \sigma_{Y_j}^2)^2} \asymp \frac{(1/n)^2}{(1/n)^2} \asymp 1.$$
Because $L$ is fixed, summing over $j$ gives
$$\sum\limits_{j=1}^L\frac{\sigma_{X_j}^2 \sigma_{Y_j}^2}{(\sigma_{X_j}^2 \beta^2 + \sigma_{Y_j}^2)^2} \asymp L \asymp 1.$$
As a result,
\begin{align*}
    V_{\mathtt{PS}}^{(1)} &= \frac{1}{\tilde{V}_2} \asymp \frac{1}{n}, \\
    V_{\mathtt{PS}}^{(2)} &= \frac{1}{\tilde{V}_2^2}\sum\limits_{j=1}^L\frac{\sigma_{X_j}^2 \sigma_{Y_j}^2}{(\sigma_{X_j}^2 \beta^2 + \sigma_{Y_j}^2)^2} \asymp \frac{1}{n^2}.
\end{align*}
By Proposition \ref{asymnorm-Lfix} and the subsequent rate analysis in Section \ref{sec:ivw-fixedL}, $V_{\mathtt{IVW}} = \Theta(n^{-1})$.
Thus,
$$V_{\mathtt{PS}}^{(1)} = \Theta(n^{-1}), \qquad V_{\mathtt{PS}}^{(2)} = \Theta(n^{-2}), \qquad V_{\mathtt{IVW}} = \Theta(n^{-1}).$$
Consequently, $$\frac{V_{\mathtt{PS}}^{(2)}}{V_{\mathtt{IVW}}} = \Theta(n^{-1}) \to 0.$$
\end{proof}

\begin{proof}[Proof of Lemma \ref{lem:exact-variance-expressions}]
Under the common variance ratio condition $\sigma_{X_j}^2 = c\sigma_{Y_j}^2$, we have $\sigma_{X_j}^2 \beta^2 + \sigma_{Y_j}^2 = (c\beta^2+1)\sigma_{Y_j}^2$. It follows that
$$\tilde{V}_2 = \sum\limits_{j=1}^L\frac{\gamma_j^2}{\sigma_{X_j}^2 \beta^2 + \sigma_{Y_j}^2} = \frac{1}{c\beta^2 + 1}\sum\limits_{j=1}^L\frac{\gamma_j^2}{\sigma_{Y_j}^2}.$$
Moreover, for each $j = 1,\dots,L$,
$$\frac{\sigma_{X_j}^2 \sigma_{Y_j}^2}{(\sigma_{X_j}^2 \beta^2 + \sigma_{Y_j}^2)^2} = \frac{c(\sigma_{Y_j}^2)^2}{(c\beta^2 + 1)^2(\sigma_{Y_j}^2)^2} = \frac{c}{(c\beta^2 + 1)^2},$$
and hence, summing over $j$,
$$\sum\limits_{j=1}^L\frac{\sigma_{X_j}^2 \sigma_{Y_j}^2}{(\sigma_{X_j}^2 \beta^2 + \sigma_{Y_j}^2)^2} = \frac{cL}{(c\beta^2 + 1)^2}.$$
Therefore,
\begin{align*}
    V_{\mathtt{PS}}^{(1)} &= \frac{1}{\tilde{V}_2} = \frac{c\beta^2 + 1}{\sum_{j=1}^L\gamma_j^2/\sigma_{Y_j}^2}, \\
    V_{\mathtt{PS}}^{(2)} &= \frac{1}{\tilde{V}_2^2}\sum\limits_{j=1}^L\frac{\sigma_{X_j}^2 \sigma_{Y_j}^2}{(\sigma_{X_j}^2 \beta^2 + \sigma_{Y_j}^2)^2} \\
    &= \frac{(c\beta^2+1)^2}{(\sum_{j=1}^L \gamma_j^2/\sigma_{Y_j}^2)^2}\frac{cL}{(c\beta^2+1)^2} \\
    &= \frac{cL}{(\sum_{j=1}^L \gamma_j^2/\sigma_{Y_j}^2)^2}.
\end{align*}
Since $c > 0$, $L \geq 1$ and $\sum_{j=1}^L\gamma_j^2 > 0$, it follows that $$V_{\mathtt{PS}}^{(2)} > 0.$$
Meanwhile, for $V_{\mathtt{IVW}}$,
\begin{align*}
    \sum\limits_{j=1}^L\frac{\gamma_j^2}{\sigma_{Y_j}^2}\frac{\beta^2\sigma_{X_j}^2+\sigma_{Y_j}^2}{\sigma_{Y_j}^2} &= (c\beta^2+1)\sum\limits_{j=1}^L\frac{\gamma_j^2}{\sigma_{Y_j}^2}, \\
    V_{\mathtt{IVW}} &= \frac{(c\beta^2+1)\sum_{j=1}^L\gamma_j^2/\sigma_{Y_j}^2}{\left(\sum_{j=1}^L\gamma_j^2/\sigma_{Y_j}^2\right)^2} = \frac{c\beta^2 + 1}{\sum_{j=1}^L\gamma_j^2/\sigma_{Y_j}^2}.
\end{align*}
Therefore, the stated expressions for $V_{\mathtt{PS}}^{(1)}$, $V_{\mathtt{PS}}^{(2)}$, and $V_{\mathtt{IVW}}$ follow.
\end{proof}

\begin{proof}[Proof of Proposition \ref{prop:diverging-L-variance-orders}]
Since $\sigma_{X_j}^2 \asymp \sigma_{Y_j}^2$ uniformly over $j=1,\dots,L$ and the structural parameter $\beta$ is fixed, it follows that $\sigma_{X_j}^2 \beta^2 + \sigma_{Y_j}^2 \asymp \sigma_{X_j}^2 \asymp \sigma_{Y_j}^2$ and so
$$\tilde{V}_2 = \sum\limits_{j=1}^L\frac{\gamma_j^2}{\sigma_{X_j}^2 \beta^2 + \sigma_{Y_j}^2} \asymp \sum\limits_{j=1}^L\frac{\gamma_j^2}{\sigma_{X_j}^2} \asymp L\kappa.$$
Moreover, for each $j = 1,\dots,L$,
$$\frac{\sigma_{X_j}^2 \sigma_{Y_j}^2}{(\sigma_{X_j}^2 \beta^2 + \sigma_{Y_j}^2)^2} \asymp \frac{\sigma_{X_j}^2 \sigma_{Y_j}^2}{\sigma_{X_j}^2 \sigma_{Y_j}^2} = 1.$$
Summing over $j$ gives
$$\sum\limits_{j=1}^L\frac{\sigma_{X_j}^2 \sigma_{Y_j}^2}{(\sigma_{X_j}^2 \beta^2 + \sigma_{Y_j}^2)^2} \asymp L.$$
As a result,
\begin{align*}
    V_{\mathtt{PS}}^{(1)} &= \frac{1}{\tilde{V}_2} \asymp \frac{1}{L\kappa}, \\
    V_{\mathtt{PS}}^{(2)} &= \frac{1}{\tilde{V}_2^2}\sum\limits_{j=1}^L\frac{\sigma_{X_j}^2 \sigma_{Y_j}^2}{(\sigma_{X_j}^2 \beta^2 + \sigma_{Y_j}^2)^2} \asymp \frac{1}{L\kappa^2}, \\
    V_{\mathtt{PS}} &\asymp \frac{1}{L\kappa}+\frac{1}{L\kappa^2} = \frac{\kappa+1}{L\kappa^2}.
\end{align*}
From Section \ref{IVW-manyL-theory},
$$\mathsf{V}_{\mathtt{IVW}} \asymp \frac{L\kappa+L}{(L\kappa+L)^2} = \frac{1}{L(\kappa+1)}.$$
$$\mathsf{V}_{\mathtt{dIVW}} \asymp \frac{L\kappa+L}{(L\kappa)^2} = \frac{\kappa+1}{L\kappa^2}.$$
The stated relative variance orders follow immediately by taking the corresponding ratios.
\end{proof}

\begin{proof}[Proof of Lemma \ref{lem:exact-variance-expressions-diverging}]
Under the common variance ratio condition $\sigma_{X_j}^2 = c\sigma_{Y_j}^2$, we have $\sigma_{X_j}^2 \beta^2 + \sigma_{Y_j}^2 = (1+c\beta^2)\sigma_{Y_j}^2$. It follows that
$$\tilde{V}_2 = \sum\limits_{j=1}^L\frac{\gamma_j^2}{\sigma_{X_j}^2 \beta^2 + \sigma_{Y_j}^2} = \frac{1}{c\beta^2 + 1}\sum\limits_{j=1}^L\frac{\gamma_j^2}{\sigma_{Y_j}^2} = \frac{cL\kappa}{1+c\beta^2}.$$
Moreover, for each $j = 1,\dots,L$,
$$\frac{\sigma_{X_j}^2 \sigma_{Y_j}^2}{(\sigma_{X_j}^2 \beta^2 + \sigma_{Y_j}^2)^2} = \frac{c(\sigma_{Y_j}^2)^2}{(c\beta^2 + 1)^2(\sigma_{Y_j}^2)^2} = \frac{c}{(c\beta^2 + 1)^2},$$
and hence, summing over $j$,
$$\sum\limits_{j=1}^L\frac{\sigma_{X_j}^2 \sigma_{Y_j}^2}{(\sigma_{X_j}^2 \beta^2 + \sigma_{Y_j}^2)^2} = \frac{cL}{(c\beta^2 + 1)^2}.$$
Therefore,
\begin{align*}
    V_{\mathtt{PS}}^{(1)} &= \frac{1}{\tilde{V}_2} = \frac{c\beta^2 + 1}{cL\kappa}, \\
    V_{\mathtt{PS}}^{(2)} &= \frac{1}{\tilde{V}_2^2}\sum\limits_{j=1}^L\frac{\sigma_{X_j}^2 \sigma_{Y_j}^2}{(\sigma_{X_j}^2 \beta^2 + \sigma_{Y_j}^2)^2} \\
    &= \frac{(1+c\beta^2)^2}{(cL\kappa)^2}\frac{cL}{(1+c\beta^2)^2} \\
    &= \frac{1}{cL\kappa^2}.
\end{align*}
Meanwhile, for $\mathsf{V}_{\mathtt{IVW}}$ and $\mathsf{V}_{\mathtt{dIVW}}$,
\begin{align*}
\mathcal{V}_1 &= \sum\limits_{j=1}^L\frac{\gamma_j^2}{\sigma_{Y_j}^2}\frac{\beta^2\sigma_{X_j}^2+\sigma_{Y_j}^2}{\sigma_{Y_j}^2} = (1+c\beta^2)\sum\limits_{j=1}^L\frac{\gamma_j^2}{\sigma_{Y_j}^2} = c(1+c\beta^2)L\kappa \\
\mathcal{V}_2 &= \sum\limits_{j=1}^L\frac{\sigma_{X_j}^2}{\sigma_{Y_j}^2}\left(1+2\beta^2\frac{\sigma_{X_j}^2}{\sigma_{Y_j}^2}\right) = c(1+2c\beta^2)L \\
\sum\limits_{j=1}^L\frac{\gamma_j^2+\sigma_{X_j}^2}{\sigma_{Y_j}^2} &= cL(\kappa+1),
\end{align*}
hence
\begin{align*}
\mathsf{V}_{\mathtt{IVW}} &= \frac{c(1+c\beta^2)L\kappa+c(1+2c\beta^2)L}{(cL(\kappa+1))^2} \\
&= \frac{(1+c\beta^2)(\kappa+1)+c\beta^2}{cL(\kappa+1)^2} \\
&= \frac{1+c\beta^2}{cL(\kappa+1)}+\frac{\beta^2}{L(\kappa+1)^2} \\
\mathsf{V}_{\mathtt{dIVW}} &= \frac{c(1+c\beta^2)L\kappa+c(1+2c\beta^2)L}{(cL\kappa)^2} \\
&= \frac{1+c\beta^2}{cL\kappa}+\frac{1+2c\beta^2}{cL\kappa^2}.
\end{align*}
\end{proof}

\end{document}